\input harvmac
\input amssym
\input epsf
%\input graphicx

% FONTS

% fraktur

\newfam\frakfam
\font\teneufm=eufm10
\font\seveneufm=eufm7
\font\fiveeufm=eufm5
\textfont\frakfam=\teneufm
\scriptfont\frakfam=\seveneufm
\scriptscriptfont\frakfam=\fiveeufm

% black board bold

\def\bb{
\font\tenmsb=msbm10
\font\sevenmsb=msbm7
\font\fivemsb=msbm5
\textfont1=\tenmsb
\scriptfont1=\sevenmsb
\scriptscriptfont1=\fivemsb
}

%\newfam\msbfam
%\font\tenmsb=msbm10
%\font\sevenmsb=msbm7
%\font\fivemsb=msbm5
%\textfont\msbfam=\tenmsb
%\scriptfont\msbfam=\sevenmsb
%\scriptscriptfont\msbfam=\fivemsb
%\def\bb{\fam\msbfam \tenmsb}

% double stroke math

\newfam\dsromfam
\font\tendsrom=dsrom10
\textfont\dsromfam=\tendsrom
\def\ds{\fam\dsromfam \tendsrom}

% bold math italics

\newfam\mbffam
\font\tenmbf=cmmib10
\font\sevenmbf=cmmib7
\font\fivembf=cmmib5
\textfont\mbffam=\tenmbf
\scriptfont\mbffam=\sevenmbf
\scriptscriptfont\mbffam=\fivembf

% bold math cal

\newfam\mbfcalfam
\font\tenmbfcal=cmbsy10
\font\sevenmbfcal=cmbsy7
\font\fivembfcal=cmbsy5
\textfont\mbfcalfam=\tenmbfcal
\scriptfont\mbfcalfam=\sevenmbfcal
\scriptscriptfont\mbfcalfam=\fivembfcal

% math script

\newfam\mscrfam
\font\tenmscr=rsfs10
\font\sevenmscr=rsfs7
\font\fivemscr=rsfs5
\textfont\mscrfam=\tenmscr
\scriptfont\mscrfam=\sevenmscr
\scriptscriptfont\mscrfam=\fivemscr
\def\scr{\fam\mscrfam \tenmscr}

% MACROS

% bras, kets, ...

% tilde, hat, bar, ...

\def\tilde{\widetilde}
\def\t{\tilde}

\def\bar{\overline}
\def\b{\bar}
\def\bsq#1{{{\b{#1}}^{\lower 2.5pt\hbox{$\scriptstyle 2$}}}}
\def\bexp#1#2{{{\b{#1}}^{\lower 2.5pt\hbox{$\scriptstyle #2$}}}}
\def\dotexp#1#2{{{#1}^{\lower 2.5pt\hbox{$\scriptstyle #2$}}}}

% basic math

\def\rt2{\sqrt{2}}
\def\half {{1 \over 2}}

\def\d{\partial}

\def\abs#1{\left|#1\right|}

% bold greek characters

\font\tenbifull=cmmib10
\font\tenbimed=cmmib7
\font\tenbismall=cmmib5
\textfont9=\tenbifull \scriptfont9=\tenbimed
\scriptscriptfont9=\tenbismall

\mathchardef\bbGamma="7000
\mathchardef\bbDelta="7001
\mathchardef\bbPhi="7002
\mathchardef\bbAlpha="7003
\mathchardef\bbXi="7004
\mathchardef\bbPi="7005
\mathchardef\bbSigma="7006
\mathchardef\bbUpsilon="7007
\mathchardef\bbTheta="7008
\mathchardef\bbPsi="7009
\mathchardef\bbOmega="700A
\mathchardef\bbalpha="710B
\mathchardef\bbbeta="710C
\mathchardef\bbgamma="710D
\mathchardef\bbdelta="710E
\mathchardef\bbepsilon="710F
\mathchardef\bbzeta="7110
\mathchardef\bbeta="7111
\mathchardef\bbtheta="7112
\mathchardef\bbiota="7113
\mathchardef\bbkappa="7114
\mathchardef\bblambda="7115
\mathchardef\bbmu="7116
\mathchardef\bbnu="7117
\mathchardef\bbxi="7118
\mathchardef\bbpi="7119
\mathchardef\bbrho="711A
\mathchardef\bbsigma="711B
\mathchardef\bbtau="711C
\mathchardef\bbupsilon="711D
\mathchardef\bbphi="711E
\mathchardef\bbchi="711F
\mathchardef\bbpsi="7120
\mathchardef\bbomega="7121
\mathchardef\bbvarepsilon="7122
\mathchardef\bbvartheta="7123
\mathchardef\bbvarpi="7124
\mathchardef\bbvarrho="7125
\mathchardef\bbvarsigma="7126
\mathchardef\bbvarphi="7127

% dotted spinor indices

\def\alphadot{{\dot\alpha}}

% bared indices

\def\ibar{\b{i}}
\def\jbar{\b{j}}
\def\kbar{\b{k}}

% bared spinors

\def\jbar{\b{j}}

% capital cal letters

\def\CA{{\cal A}}

\def\CJ{{\cal J}}

\def\CM{{\cal M}}
\def\CN{{\cal N}}
\def\CO{{\cal O}}

\def\CR{{\cal R}}
\def\CS{{\cal S}}

% double stroke symbols: unit matrix, reals, complex, quaternions, integers, natural numbers

\def\1{{\ds 1}}

\def\C{\hbox{$\bb C$}}

\def\P{\hbox{$\bb P$}}

% miscellaneous objects

\def\ep{\varepsilon}

\def\CM{{\cal M}}

\def\K3{{\bf K3}}
\def\journal#1&#2(#3){\unskip, \sl #1\ \bf #2 \rm(19#3) }
\def\andjournal#1&#2(#3){\sl #1~\bf #2 \rm (19#3) }

\def\bar{\overline}

\def\tilde{\widetilde}

\def\frac#1#2{{#1\over#2}}

\def\half{\frac12}

\def\d{\partial}

\def\inbar{\,\vrule height1.5ex width.4pt depth0pt}
\def\IC{\relax\hbox{$\inbar\kern-.3em{\rm C}$}}
\def\IR{\relax{\rm I\kern-.18em R}}
\def\IP{\relax{\rm I\kern-.18em P}}

%
%%%%%%%%%%%%%%%%%%%%%%%%%%%%%%%%%%%%
%

%
\catcode`\@=11
\def\slash#1{\mathord{\mathpalette\c@ncel{#1}}}
\overfullrule=0pt

\def\underrel#1\over#2{\mathrel{\mathop{\kern\z@#1}\limits_{#2}}}

\catcode`\@=12

%%%%%%%%%%%%%%%%%%%%%%%%%%%%%%%%%%%%%%%%%%%%%%%%%%%%%%%%%%%%%%

%

\def\unit{\relax{\rm 1\kern-.26em I}}
\def\nada{\relax{\rm 0\kern-.30em l}}
\def\tilde{\widetilde}
\def\t{\tilde}
\def\alphadot{{\dot \alpha}}

%\def\Omega{\rho,\sigma,\nu  }

%% MACROS
\noblackbox
\def\IL{\relax{\rm I\kern-.18em L}}
\def\IH{\relax{\rm I\kern-.18em H}}
\def\IR{\relax{\rm I\kern-.18em R}}
\def\IC{\relax\hbox{$\inbar\kern-.3em{\rm C}$}}
\def\IZ{\relax\ifmmode\mathchoice
{\hbox{\cmss Z\kern-.4em Z}}{\hbox{\cmss Z\kern-.4em Z}} {\lower.9pt\hbox{\cmsss Z\kern-.4em Z}}
{\lower1.2pt\hbox{\cmsss Z\kern-.4em Z}}\else{\cmss Z\kern-.4em Z}\fi}
\def\CM {{\cal M}}
\def\CN {{\cal N}}
\def\CR {{\cal R}}

\def\CJ {{\cal J}}

\def\CO {{\cal O}}

\def\CS {{\cal S}}
\def\CA{{\cal A}}

%% MORE MACROS
\def\CM {{\cal M}}
\def\CN {{\cal N}}

\def\CO {{\cal O}}
\def\ind{{^{(a)}}}

\def\CS {{\cal S }}

\font\manual=manfnt \def\dbend{\lower3.5pt\hbox{\manual\char127}}

\def\IZ{\relax\ifmmode\mathchoice
{\hbox{\cmss Z\kern-.4em Z}}{\hbox{\cmss Z\kern-.4em Z}} {\lower.9pt\hbox{\cmsss Z\kern-.4em Z}}
{\lower1.2pt\hbox{\cmsss Z\kern-.4em Z}}\else{\cmss Z\kern-.4em Z}\fi}
\def\half {{1\over 2}}

\def\alphadot{{\dot \alpha}}

\def\ibar{{\bar i}}
\def\jbar{{\bar j}}
\def\kbar{{\bar k}}

\def\bar{\overline}
\def\CS{{\cal S}}

\def\pa{\partial}

\def\rt2{\sqrt{2}}
\def\irt2{{1\over\sqrt{2}}}

\def\t{\tilde}

%  \slashchar puts a slash through a character to represent contraction
%  with Dirac matrices. Use \not instead for negation of relations, and use
%  \hbar for hbar.
\def\slashchar#1{\setbox0=\hbox{$#1$}           % set a box for #1
   \dimen0=\wd0                                 % and get its size
   \setbox1=\hbox{/} \dimen1=\wd1               % get size of /
   \ifdim\dimen0>\dimen1                        % #1 is bigger
      \rlap{\hbox to \dimen0{\hfil/\hfil}}      % so center / in box
      #1                                        % and print #1
   \else                                        % / is bigger
      \rlap{\hbox to \dimen1{\hfil$#1$\hfil}}   % so center #1
      /                                         % and print /
   \fi}

\def\foursqr#1#2{{\vcenter{\vbox{
    \hrule height.#2pt
    \hbox{\vrule width.#2pt height#1pt \kern#1pt
    \vrule width.#2pt}
    \hrule height.#2pt
    \hrule height.#2pt
    \hbox{\vrule width.#2pt height#1pt \kern#1pt
    \vrule width.#2pt}
    \hrule height.#2pt
        \hrule height.#2pt
    \hbox{\vrule width.#2pt height#1pt \kern#1pt
    \vrule width.#2pt}
    \hrule height.#2pt
        \hrule height.#2pt
    \hbox{\vrule width.#2pt height#1pt \kern#1pt
    \vrule width.#2pt}
    \hrule height.#2pt}}}}
\def\psqr#1#2{{\vcenter{\vbox{\hrule height.#2pt
    \hbox{\vrule width.#2pt height#1pt \kern#1pt
    \vrule width.#2pt}
    \hrule height.#2pt \hrule height.#2pt
    \hbox{\vrule width.#2pt height#1pt \kern#1pt
    \vrule width.#2pt}
    \hrule height.#2pt}}}}
\def\sqr#1#2{{\vcenter{\vbox{\hrule height.#2pt
    \hbox{\vrule width.#2pt height#1pt \kern#1pt
    \vrule width.#2pt}
    \hrule height.#2pt}}}}
\def\square{\mathchoice\sqr65\sqr65\sqr{2.1}3\sqr{1.5}3}

\def\figin{\epsfcheck\figin}\def\figins{\epsfcheck\figins}
\def\epsfcheck{\ifx\epsfbox\UnDeFiNeD
\message{(NO epsf.tex, FIGURES WILL BE IGNORED)}
\gdef\figin##1{\vskip2in}\gdef\figins##1{\hskip.5in}% blank space instead
\else\message{(FIGURES WILL BE INCLUDED)}%
\gdef\figin##1{##1}\gdef\figins##1{##1}\fi}
\def\DefWarn#1{}
\def\figinsert{\goodbreak\midinsert}
\def\ifig#1#2#3{\DefWarn#1\xdef#1{fig.~\the\figno}
\writedef{#1\leftbracket fig.\noexpand~\the\figno}%
\figinsert\figin{\centerline{#3}}\medskip\centerline{\vbox{\baselineskip12pt \advance\hsize by
-1truein\noindent\footnotefont{\bf Fig.~\the\figno:\ } \it#2}}
\bigskip\endinsert\global\advance\figno by1}

%%%%%%%%%%%%%%%%%%%%%%%%%%%%%%%%%%%%%%%%%%%%%%%%%%%%%%%%%%%%%%
% new defs:

\def\ibar{{\bar i}}
\def\jbar{{\bar j}}
\def\kbar{{\bar k}}

%\ElvangGK
\lref\ElvangGK{
  H.~Elvang and B.~Wecht,
  ``Semi-Direct Gauge Mediation with the 4-1 Model,''
  JHEP {\bf 0906}, 026 (2009)
  [arXiv:0904.4431 [hep-ph]].
  %%CITATION = JHEPA,0906,026;%%
}

%\PoppitzFH
\lref\PoppitzFH{
  E.~Poppitz and S.~P.~Trivedi,
  ``Some examples of chiral moduli spaces and dynamical supersymmetry
  breaking,''
  Phys.\ Lett.\  B {\bf 365}, 125 (1996)
  [arXiv:hep-th/9507169].
  %%CITATION = PHLTA,B365,125;%%
}

%\FischlerZK
\lref\FischlerZK{
  W.~Fischler, H.~P.~Nilles, J.~Polchinski, S.~Raby and L.~Susskind,
  ``Vanishing Renormalization Of The D Term In Supersymmetric U(1) Theories,''
  Phys.\ Rev.\ Lett.\  {\bf 47}, 757 (1981).
  %%CITATION = PRLTA,47,757;%%
}

%\FerraraPZ
\lref\FerraraPZ{
  S.~Ferrara and B.~Zumino,
  ``Transformation Properties Of The Supercurrent,''
  Nucl.\ Phys.\  B {\bf 87}, 207 (1975).
  %%CITATION = NUPHA,B87,207;%%
}

%\IntriligatorCP
\lref\IntriligatorCP{
  K.~A.~Intriligator and N.~Seiberg,
  ``Lectures on Supersymmetry Breaking,''
  Class.\ Quant.\ Grav.\  {\bf 24}, S741 (2007)
  [arXiv:hep-ph/0702069].
  %%CITATION = CQGRD,24,S741;%%
}

%\DineTA
\lref\DineTA{
  M.~Dine,
  ``Fields, Strings and Duality: TASI 96,''
 eds. C.~Efthimiou and B. Greene (World Scientific, Singapore, 1997).
}

%\FayetJB
\lref\FayetJB{
  P.~Fayet and J.~Iliopoulos,
  ``Spontaneously Broken Supergauge Symmetries and Goldstone Spinors,''
  Phys.\ Lett.\  B {\bf 51}, 461 (1974).
  %%CITATION = PHLTA,B51,461;%%
}

%\HullPQ
\lref\HullPQ{
  C.~M.~Hull, A.~Karlhede, U.~Lindstrom and M.~Rocek,
  ``Nonlinear Sigma Models And Their Gauging In And Out Of Superspace,''
  Nucl.\ Phys.\  B {\bf 266}, 1 (1986).
  %%CITATION = NUPHA,B266,1;%%
}

%\BaggerQH
\lref\BaggerQH{
  J.~Bagger and J.~Wess,
  ``Supersymmetry and supergravity,''
  %%CITATION = JHU-TIPAC-9009;%%
}

%\KomargodskiRB
\lref\KomargodskiRB{
  Z.~Komargodski and N.~Seiberg,
  ``Comments on Supercurrent Multiplets, Supersymmetric Field Theories and
  Supergravity,''
  arXiv:1002.2228 [hep-th].
  %%CITATION = ARXIV:1002.2228;%%
}

%\AffleckXZ
\lref\AffleckXZ{
  I.~Affleck, M.~Dine and N.~Seiberg,
  ``Dynamical Supersymmetry Breaking In Four-Dimensions And Its
  Phenomenological Implications,''
  Nucl.\ Phys.\  B {\bf 256}, 557 (1985).
  %%CITATION = NUPHA,B256,557;%%
}

%\IntriligatorPU
\lref\IntriligatorPU{
  K.~A.~Intriligator and S.~D.~Thomas,
  ``Dynamical Supersymmetry Breaking on Quantum Moduli Spaces,''
  Nucl.\ Phys.\  B {\bf 473}, 121 (1996)
  [arXiv:hep-th/9603158].
  %%CITATION = NUPHA,B473,121;%%
}

%\KomargodskiPC
\lref\KomargodskiPC{
  Z.~Komargodski and N.~Seiberg,
  ``Comments on the Fayet-Iliopoulos Term in Field Theory and Supergravity,''
  JHEP {\bf 0906}, 007 (2009)
  [arXiv:0904.1159 [hep-th]].
  %%CITATION = JHEPA,0906,007;%%
}

%\BaggerFN
\lref\BaggerFN{
  J.~Bagger and E.~Witten,
  ``The Gauge Invariant Supersymmetric Nonlinear Sigma Model,''
  Phys.\ Lett.\  B {\bf 118}, 103 (1982).
  %%CITATION = PHLTA,B118,103;%%
}

%\GreenDA
\lref\GreenDA{
  D.~Green, Z.~Komargodski, N.~Seiberg, Y.~Tachikawa and B.~Wecht,
  ``Exactly Marginal Deformations and Global Symmetries,''
  arXiv:1005.3546 [hep-th].
  %%CITATION = ARXIV:1005.3546;%%
}

%\CallanZE
\lref\CallanZE{
  C.~G.~.~Callan, S.~R.~Coleman and R.~Jackiw,
  ``A New improved energy - momentum tensor,''
  Annals Phys.\  {\bf 59}, 42 (1970).
  %%CITATION = APNYA,59,42;%%
}

%\BandoCC
\lref\BandoCC{
  M.~Bando, T.~Kuramoto, T.~Maskawa and S.~Uehara,
  ``Nonlinear Realization In Supersymmetric Theories,''
  Prog.\ Theor.\ Phys.\  {\bf 72}, 313 (1984).
  %%CITATION = PTPKA,72,313;%%
}

%\BandoFN
\lref\BandoFN{
  M.~Bando, T.~Kuramoto, T.~Maskawa and S.~Uehara,
  ``Nonlinear Realization In Supersymmetric Theories. 2,''
  Prog.\ Theor.\ Phys.\  {\bf 72}, 1207 (1984).
  %%CITATION = PTPKA,72,1207;%%
}

%\KugoMA
\lref\KugoMA{
  T.~Kugo, I.~Ojima and T.~Yanagida,
  ``Superpotential Symmetries And Pseudonambu-Goldstone Supermultiplets,''
  Phys.\ Lett.\  B {\bf 135}, 402 (1984).
  %%CITATION = PHLTA,B135,402;%%
}

%\LercheQA
\lref\LercheQA{
  W.~Lerche,
  ``On Goldstone Fields In Supersymmetric Theories,''
  Nucl.\ Phys.\  B {\bf 238}, 582 (1984).
  %%CITATION = NUPHA,B238,582;%%
}

%\ShoreTF
\lref\shoreb{
  G.~M.~Shore,
  ``The Supersymmetric Higgs Mechanism: Quartet Decoupling And Nondoubled
  Goldstone Bosons,''
  Annals Phys.\  {\bf 168}, 46 (1986).
  %%CITATION = APNYA,168,46;%%
}

%\ShoreBH
\lref\shorea{
  G.~M.~Shore,
  ``Supersymmetric Higgs Mechanism With Nondoubled Goldstone Bosons,''
  Nucl.\ Phys.\  B {\bf 248}, 123 (1984).
  %%CITATION = NUPHA,B248,123;%%
}

%\GregoireJR
\lref\GregoireJR{
  T.~Gregoire, R.~Rattazzi and C.~A.~Scrucca,
  ``D-type supersymmetry breaking and brane-to-brane gravity mediation,''
  Phys.\ Lett.\  B {\bf 624}, 260 (2005)
  [arXiv:hep-ph/0505126].
  %%CITATION = PHLTA,B624,260;%%
}

%\WittenHU
\lref\WittenHU{
  E.~Witten and J.~Bagger,
  ``Quantization Of Newton's Constant In Certain Supergravity Theories,''
  Phys.\ Lett.\  B {\bf 115}, 202 (1982).
  %%CITATION = PHLTA,B115,202;%%
}

%\SeibergBZ
\lref\SeibergBZ{
  N.~Seiberg,
  ``Exact Results On The Space Of Vacua Of Four-Dimensional Susy Gauge
  Theories,''
  Phys.\ Rev.\  D {\bf 49}, 6857 (1994)
  [arXiv:hep-th/9402044].
  %%CITATION = PHRVA,D49,6857;%%
}

%\AffleckMK
\lref\AffleckMK{
  I.~Affleck, M.~Dine and N.~Seiberg,
  ``Dynamical Supersymmetry Breaking In Supersymmetric QCD,''
  Nucl.\ Phys.\  B {\bf 241}, 493 (1984).
  %%CITATION = NUPHA,B241,493;%%
}

%\DavisMZ
\lref\DavisMZ{
  A.~C.~Davis, M.~Dine and N.~Seiberg,
  ``The Massless Limit Of Supersymmetric QCD,''
  Phys.\ Lett.\  B {\bf 125}, 487 (1983).
  %%CITATION = PHLTA,B125,487;%%
}

%\CarpenterTZ
\lref\CarpenterTZ{
  L.~M.~Carpenter, P.~J.~Fox and D.~E.~Kaplan,
  ``The NMSSM, anomaly mediation and a Dirac Bino,''
  arXiv:hep-ph/0503093.
  %%CITATION = HEP-PH/0503093;%%
}

%\DineAG
\lref\DineAG{
  M.~Dine, A.~E.~Nelson, Y.~Nir and Y.~Shirman,
  ``New tools for low-energy dynamical supersymmetry breaking,''
  Phys.\ Rev.\  D {\bf 53}, 2658 (1996)
  [arXiv:hep-ph/9507378].
  %%CITATION = PHRVA,D53,2658;%%
}

%\IntriligatorFK
\lref\IntriligatorFK{
  K.~A.~Intriligator and S.~D.~Thomas,
  ``Dual descriptions of supersymmetry breaking,''
  arXiv:hep-th/9608046.
  %%CITATION = HEP-TH/9608046;%%
}

%\ShadmiJY
\lref\ShadmiJY{
  Y.~Shadmi and Y.~Shirman,
  ``Dynamical supersymmetry breaking,''
  Rev.\ Mod.\ Phys.\  {\bf 72}, 25 (2000)
  [arXiv:hep-th/9907225].
  %%CITATION = RMPHA,72,25;%%
}

%\SeibergPQ
\lref\SeibergPQ{
  N.~Seiberg,
  ``Electric - magnetic duality in supersymmetric nonabelian gauge theories,''
  Nucl.\ Phys.\  B {\bf 435}, 129 (1995)
  [arXiv:hep-th/9411149].
  %%CITATION = NUPHA,B435,129;%%
}

%\IntriligatorDD
\lref\IntriligatorDD{
  K.~A.~Intriligator, N.~Seiberg and D.~Shih,
  ``Dynamical SUSY breaking in meta-stable vacua,''
  JHEP {\bf 0604}, 021 (2006)
  [arXiv:hep-th/0602239].
  %%CITATION = JHEPA,0604,021;%%
}

%\GiveonNE
\lref\GiveonNE{
  A.~Giveon, A.~Katz, Z.~Komargodski and D.~Shih,
  ``Dynamical SUSY and R-symmetry breaking in SQCD with massive and massless
  flavors,''
  JHEP {\bf 0810}, 092 (2008)
  [arXiv:0808.2901 [hep-th]].
  %%CITATION = JHEPA,0810,092;%%
}
%\GiveonWP
\lref\GiveonWP{
  A.~Giveon, A.~Katz and Z.~Komargodski,
  ``On SQCD with massive and massless flavors,''
  JHEP {\bf 0806}, 003 (2008)
  [arXiv:0804.1805 [hep-th]].
  %%CITATION = JHEPA,0806,003;%%
}

%\AmaritiUZ
\lref\AmaritiUZ{
  A.~Amariti and A.~Mariotti,
  ``Two Loop R-Symmetry Breaking,''
  JHEP {\bf 0907}, 071 (2009)
  [arXiv:0812.3633 [hep-th]].
  %%CITATION = JHEPA,0907,071;%%
}

%\IntriligatorBE
\lref\IntriligatorBE{
  K.~Intriligator and M.~Sudano,
  ``General Gauge Mediation with Gauge Messengers,''
  arXiv:1001.5443 [hep-ph].
  %%CITATION = ARXIV:1001.5443;%%
}

%\BuicanVV
\lref\BuicanVV{
  M.~Buican and Z.~Komargodski,
  ``Soft Terms from Broken Symmetries,''
  JHEP {\bf 1002}, 005 (2010)
  [arXiv:0909.4824 [hep-ph]].
  %%CITATION = JHEPA,1002,005;%%
}

%\IntriligatorFR
\lref\IntriligatorFR{
  K.~A.~Intriligator and M.~Sudano,
  ``Comments on General Gauge Mediation,''
  JHEP {\bf 0811}, 008 (2008)
  [arXiv:0807.3942 [hep-ph]].
  %%CITATION = JHEPA,0811,008;%%
}

%\GorbatovQA
\lref\GorbatovQA{
  E.~Gorbatov and M.~Sudano,
  ``Sparticle Masses in Higgsed Gauge Mediation,''
  JHEP {\bf 0810}, 066 (2008)
  [arXiv:0802.0555 [hep-ph]].
  %%CITATION = JHEPA,0810,066;%%
}

%\IzawaPK
\lref\IzawaPK{
  K.~I.~Izawa and T.~Yanagida,
  ``Dynamical Supersymmetry Breaking in Vector-like Gauge Theories,''
  Prog.\ Theor.\ Phys.\  {\bf 95}, 829 (1996)
  [arXiv:hep-th/9602180].
  %%CITATION = PTPKA,95,829;%%
}

%\IntriligatorPU
\lref\IntriligatorPU{
  K.~A.~Intriligator and S.~D.~Thomas,
  ``Dynamical Supersymmetry Breaking on Quantum Moduli Spaces,''
  Nucl.\ Phys.\  B {\bf 473}, 121 (1996)
  [arXiv:hep-th/9603158].
  %%CITATION = NUPHA,B473,121;%%
}

%\DumitrescuHA
\lref\DumitrescuHA{
  T.~T.~Dumitrescu, Z.~Komargodski, N.~Seiberg and D.~Shih,
  ``General Messenger Gauge Mediation,''
  JHEP {\bf 1005}, 096 (2010)
  [arXiv:1003.2661 [hep-ph]].
  %%CITATION = JHEPA,1005,096;%%
}

%\BuicanWS
\lref\BuicanWS{
  M.~Buican, P.~Meade, N.~Seiberg and D.~Shih,
  ``Exploring General Gauge Mediation,''
  JHEP {\bf 0903}, 016 (2009)
  [arXiv:0812.3668 [hep-ph]].
  %%CITATION = JHEPA,0903,016;%%
}

%\MeadeWD
\lref\MeadeWD{
  P.~Meade, N.~Seiberg and D.~Shih,
  ``General Gauge Mediation,''
  Prog.\ Theor.\ Phys.\ Suppl.\  {\bf 177}, 143 (2009)
  [arXiv:0801.3278 [hep-ph]].
  %%CITATION = PTPSA,177,143;%%
}

%\BenakliGI
\lref\BenakliGI{
  K.~Benakli and M.~D.~Goodsell,
  ``Dirac Gauginos, Gauge Mediation and Unification,''
  arXiv:1003.4957 [hep-ph].
  %%CITATION = ARXIV:1003.4957;%%
}

%\BenakliPG
\lref\BenakliPG{
  K.~Benakli and M.~D.~Goodsell,
  ``Dirac Gauginos in General Gauge Mediation,''
  Nucl.\ Phys.\  B {\bf 816}, 185 (2009)
  [arXiv:0811.4409 [hep-ph]].
  %%CITATION = NUPHA,B816,185;%%
}

%\RayWK
\lref\RayWK{
  S.~Ray,
  ``Some properties of meta-stable supersymmetry-breaking vacua in Wess-Zumino
  models,''
  Phys.\ Lett.\  B {\bf 642}, 137 (2006)
  [arXiv:hep-th/0607172].
  %%CITATION = PHLTA,B642,137;%%
}

%\AuzziKV
\lref\AuzziKV{
  R.~Auzzi and E.~Rabinovici,
  ``On metastable vacua in perturbed N=2 theories,''
  arXiv:1006.0637 [hep-th].
  %%CITATION = ARXIV:1006.0637;%%
}

%\OoguriIU
\lref\OoguriIU{
  H.~Ooguri, Y.~Ookouchi and C.~S.~Park,
  ``Metastable Vacua in Perturbed Seiberg-Witten Theories,''
  Adv.\ Theor.\ Math.\ Phys.\  {\bf 12}, 405 (2008)
  [arXiv:0704.3613 [hep-th]].
  %%CITATION = 00203,12,405;%%
}

%\MarsanoMT
\lref\MarsanoMT{
  J.~Marsano, H.~Ooguri, Y.~Ookouchi and C.~S.~Park,
  ``Metastable Vacua in Perturbed Seiberg-Witten Theories, Part 2:
  Fayet-Iliopoulos Terms and K\'ahler Normal Coordinates,''
  Nucl.\ Phys.\  B {\bf 798}, 17 (2008)
  [arXiv:0712.3305 [hep-th]].
  %%CITATION = NUPHA,B798,17;%%
}

%\AntoniadisNJ
\lref\AntoniadisNJ{
  I.~Antoniadis and M.~Buican,
  ``Goldstinos, Supercurrents and Metastable SUSY Breaking in N=2
  Supersymmetric Gauge Theories,''
  arXiv:1005.3012 [hep-th].
  %%CITATION = ARXIV:1005.3012;%%
}

%\DienesTD
\lref\DienesTD{
  K.~R.~Dienes and B.~Thomas,
  ``On the Inconsistency of Fayet-Iliopoulos Terms in Supergravity Theories,''
  Phys.\ Rev.\  D {\bf 81}, 065023 (2010)
  [arXiv:0911.0677 [hep-th]].
  %%CITATION = PHRVA,D81,065023;%%
}

%\KuzenkoYM
\lref\KuzenkoYM{
  S.~M.~Kuzenko,
  ``The Fayet-Iliopoulos term and nonlinear self-duality,''
  Phys.\ Rev.\  D {\bf 81}, 085036 (2010)
  [arXiv:0911.5190 [hep-th]].
  %%CITATION = PHRVA,D81,085036;%%
}

%\PoppitzVD
\lref\PoppitzVD{
  E.~Poppitz and S.~P.~Trivedi,
  ``Dynamical supersymmetry breaking,''
  Ann.\ Rev.\ Nucl.\ Part.\ Sci.\  {\bf 48}, 307 (1998)
  [arXiv:hep-th/9803107].
  %%CITATION = ARNUA,48,307;%%
}

%\yutyuich
\lref\yutyuich{
  Y.~Nakai and Y.~Ookouchi,
  To Appear.
  %%CITATION = ARNUA,48,307;%%
}

%\MatosXV
\lref\matos{
  L.~F.~Matos,
  ``Some examples of F and D-term SUSY breaking models,''
  arXiv:0910.0451 [hep-ph].
  %%CITATION = ARXIV:0910.0451;%%
}

%\SeibergQJ
\lref\SeibergQJ{
  N.~Seiberg, T.~Volansky and B.~Wecht,
  ``Semi-direct Gauge Mediation,''
  JHEP {\bf 0811}, 004 (2008)
  [arXiv:0809.4437 [hep-ph]].
  %%CITATION = JHEPA,0811,004;%%
}

%\MagroAJ
\lref\MagroAJ{
  M.~Magro, I.~Sachs and S.~Wolf,
  ``Superfield Noether procedure,''
  Annals Phys.\  {\bf 298}, 123 (2002)
  [arXiv:hep-th/0110131].
  %%CITATION = APNYA,298,123;%%
}

% DRAFTMODE

%\draftmode

% TITLE PAGE

\rightline{IPMU 10-0129}
\vskip-2pt
\rightline{PUPT-2346}

\Title{\vbox{\baselineskip5pt \hbox{}}}
{\vbox{\centerline{Global Symmetries and $D$-Terms}
\vskip5pt
\centerline{in Supersymmetric Field Theories}}}

\vskip-10pt
\centerline{Thomas T. Dumitrescu,$^1$ Zohar Komargodski,$^2$ and Matthew Sudano$^3$}

\vskip14pt

\centerline{$^1${\it Department of Physics, Princeton University, Princeton, NJ 08544, USA}}

\vskip4pt

\centerline{$^2${\it  School of Natural Sciences, Institute for Advanced Study, Princeton, NJ 08540, USA}}

\vskip4pt

\centerline{$^3${\it  Institute for the Physics and Mathematics of the Universe, }}
\vskip-2pt
\centerline{\it University of Tokyo, Kashiwa, Chiba 277-8568, Japan}

\vskip1cm

\noindent We study the
role of~$D$-terms in supersymmetry (SUSY) breaking. By carefully analyzing the SUSY multiplets containing various conserved currents in theories with global symmetries, we obtain a number of constraints on the renormalization group flow in supersymmetric field theories. Under broad assumptions, these results imply that there are no SUSY-breaking vacua, not even metastable ones, with parametrically large~$D$-terms. This explains the absence of such~$D$-terms in models of dynamical SUSY-breaking. There is, however, a rich class of calculable models which generate comparable~$D$-terms and~$F$-terms through a variety of non-perturbative effects; these~$D$-terms can be non-abelian. We give several explicit examples of such models, one of which is a new calculable limit of the~$3$-$2$ model.

\vskip2.5cm

\Date{July 2010}

%%%%%%%%%%%%%%%%%%%%%%%%%%%%%%%%%%%%%%%%%%%%%%%%%%%

\newsec{Introduction}

In local quantum field theories, the basic objects of interest are
well-defined local operators. Conserved currents are an important
class of such local operators. In theories with continuous symmetries, their existence is guaranteed by the Noether theorem.

One usually considers two types of conserved currents: global
symmetry currents~$j_\mu^{(a)}$ (we will use indices~$a, b,
\ldots$ to label different global symmetries) and spacetime
symmetry currents, such as the energy-momentum tensor~$T_{\mu\nu}$: \eqn\currcons{\d^\mu j_\mu^{(a)} = 0~, \qquad \d^\mu
T_{\mu\nu}=0~.} Given a microscopic description of the theory, the Noether theorem tells us how to express~$j_\mu^{(a)}$ and~$T_{\mu\nu}$ in terms of elementary
fields. These expressions are not unique; they are only defined up
to improvement terms -- operators which are
automatically conserved and do not affect the charges.

The improvement terms can usually be chosen in such a way that the
currents satisfy the following basic conditions:

\medskip

\item{1.)} They are gauge-invariant local operators.

\item{2.)} They are globally well-defined, even if the configuration space of the theory is non-trivial.

\item{3.)} They satisfy the current algebra of symmetries.

\medskip

\noindent In particular, condition $(3)$ means that the energy-momentum tensor is invariant under global symmetry
transformations, and that the global symmetry currents satisfy the
usual current algebra. We will refer to $(1)-(3)$ as {\it
consistency conditions}.  Note that the conservation
equations~\currcons\ and the consistency conditions~$(1)-(3)$ are
statements about operators. As such, they are insensitive to
possible symmetry breaking by the vacuum.

In four-dimensional theories with~$\CN=1$ supersymmetry, there are two additional ingredients. First, there is a conserved supercurrent~$S_{\mu\alpha}$~$\,\left(\d^\mu S_{\mu\alpha} = 0\right)$; it too is only defined up to improvement terms. Second, gauge-invariant local operators are embedded in supermultiplets. In particular, the supercurrent, the energy-momentum tensor and the global symmetry currents should all be embedded in supermultiplets. This will play a crucial role in our analysis.

Global symmetry currents are conventionally embedded in a real multiplet~$\CJ^{(a)}$ satisfying~$\bar D^2\CJ\ind=0$. In components, it takes the form
\eqn\complinear{\CJ\ind=J\ind+i\theta j\ind-i\bar\theta\bar j\ind-\theta\sigma^\mu\bar\theta j^{(a)}_\mu+\cdots~.}
We see that the conserved current~$j_\mu^{(a)}$ is accompanied by spin-$0$ and spin-$\half$ operators~$J\ind$ and~$j^{(a)}_\alpha$.

The supercurrent and the energy-momentum tensor are conventionally
embedded in the Ferrara-Zumino (FZ) multiplet \FerraraPZ.
Schematically, it takes the form
\eqn\FZmul{\CJ_\mu^{FZ} \sim j_\mu^{FZ}+\theta^\alpha S_{\mu\alpha}
+\bar\theta_{\alphadot} \bar
S_\mu^{\alphadot} + \left(\theta\sigma^\nu\bar\theta\right) T_{\mu\nu} + \cdots~.}
The FZ-multiplet is real and
satisfies the conservation equation~$\bar
D^{\alphadot}\CJ_{\alpha\alphadot}^{FZ}=D_\alpha X$, where $X$ is
a chiral superfield.\foot{Our convention for switching between
vectors and bi-spinors is
$$\ell_{\alpha\alphadot}=-2\sigma^\mu_{\alpha\alphadot}\ell_\mu~,\qquad
\ell_\mu={1\over
4}\bar\sigma_\mu^{\alphadot\alpha}\ell_{\alpha\alphadot}~.$$ Any
unstated conventions are those of Wess and Bagger~\BaggerQH.} This
contains the conservation equations for~$S_{\mu\alpha}$ and~$T_{\mu\nu}$, as well as the statement that $T_{\mu\nu}$ is symmetric.
Moreover, we see that~$S_{\mu\alpha}$ and~$T_{\mu\nu}$ are in the same
multiplet as a vector operator~$j_\mu^{FZ}$, which in
general need not be conserved.\foot{The multiplet also contains a complex scalar (the bottom component of~$X$), which will not appear in our analysis.}

We would like to understand whether the consistency conditions~$(1)-(3)$ are satisfied by the operators in the supermultiplets~$\CJ\ind$ and~$\CJ_{\alpha\alphadot}^{FZ}$. This is a subtle point which needs some clarification. It is usually possible choose improvement terms such that the operators~$j_\mu^{(a)}$, $S_{\mu\alpha}$ and~$T_{\mu\nu}$ satisfy the consistency conditions. However, a problem could arise when trying to embed these operators in the SUSY multiplets discussed above. For example, to embed $T_{\mu\nu}$ in the FZ-multiplet, we might be forced to pick the improvement terms in such a way that some of the consistency conditions are violated. Another possibility is that the operators~$j_\mu^{(a)}$,~$S_{\mu\alpha}$ and~$T_{\mu\nu}$ in the multiplets~$\CJ\ind$ and~$\CJ_{\alpha\alphadot}^{FZ}$ do satisfy the consistency conditions, but that the lower-spin operators in these multiplets do not.

Whenever such a situation arises, we will say that there is a
{\it clash} between the existence of the usual supersymmetry
multiplets~\complinear\ and~\FZmul, and the consistency
conditions~$(1)-(3)$. We emphasize that this clash refers to the familiar, well-studied SUSY multiplets~\complinear\ and~\FZmul. There may be other (generally larger) supersymmetry multiplets that do satisfy the consistency conditions. Our results do not depend on the existence of these other multiplets.

For the FZ-multiplet, the consistency conditions~$(1)$ and~$(2)$ were studied in~\refs{\KomargodskiPC,\KomargodskiRB}. This has led to new results about rigid supersymmetric
theories and supergravity. A simple example, in which a clash of the type described above occurs, is a free~$U(1)$ gauge theory with a Fayet-Iliopoulos~(FI) term:
\eqn\thfi{{\scr L} = {1\over
4g^2}\int d^2\theta \, W_\alpha^2+ {\rm c.c.} +\xi\int d^4\theta \, V~.}
The
FZ-multiplet of this theory is given by
\eqn\Fzmulfi{\CJ^{FZ}_{\alpha\alphadot}=-{4\over g^2}W_\alpha\bar
W_{\alphadot}-{2\over 3}\xi[D_\alpha,\bar D_{\alphadot}]V~.} We
see that a gauge transformation $V \rightarrow
V+i(\Lambda-\b\Lambda)$ does not leave~\Fzmulfi\ invariant.  In
other words, there is a clash between the existence of the
FZ-multiplet and gauge invariance. This does not render the theory
inconsistent. In components, the gauge non-invariance takes the
form of an improvement term for both the supercurrent and the
energy-momentum tensor. Thus, the supersymmetry charges and the
momentum operators are gauge-invariant. Furthermore, there is a
larger supersymmetry multiplet, the $\CS$-multiplet
described in~\refs{\MagroAJ,\KomargodskiRB}, which contains a conserved supercurrent and a conserved energy-momentum tensor but is also gauge-invariant.\foot{The simple theory in~\thfi\ has an $R$-symmetry so that there is another well-defined supercurrent multiplet~\refs{\DienesTD,\KuzenkoYM}. This is the~$\CR$-multiplet reviewed, for example, in~\KomargodskiRB.}

The importance of this clash is that it leads to a
non-renormalization theorem: if the FZ-multiplet~\Fzmulfi\ is
gauge-invariant in the UV, then this must continue to hold at every
energy scale. In the example above, this means that if $\xi = 0$
in the UV then a non-zero~$\xi$ cannot be generated under renormalization group~(RG) flow.
In the case of~\thfi\ this statement is trivial because the theory
is free. However, the non-renonormalization theorem holds even in the
presence of matter and strong gauge dynamics: if a theory has a
gauge-invariant FZ-multiplet in the UV, then a FI-term cannot be
generated, perturbatively or non-perturbatively, for any~$U(1)$ gauge
group. This applies even to~$U(1)$ gauge groups that emerge in the IR from
the dynamics of the theory. These observations partly explain the absence of
large~$D$-terms in models of dynamical SUSY-breaking.

In this paper we study rigid supersymmetric theories with global
symmetries. We analyze the consistency conditions~$(1)-(3)$ for
the global current multiplets~$\CJ\ind$ and for the
FZ-multiplet~$\CJ^{FZ}_{\alpha\alphadot}$.
This leads to new constraints on the RG-flow of rigid supersymmetric theories. Moreover, it allows us to make general statements about the role of~$D$-terms in supersymmetry breaking.

In section~2 we consider supersymmetric sigma
models and their global symmetries in detail. These models often arise as the low-energy limit of interesting supersymmetric field theories. We begin by discussing the conditions for the current multiplets~$\CJ\ind$ to be globally well-defined, and we analyze the current algebra they satisfy. We then turn to the FZ-multiplet for sigma models. After briefly reviewing when the FZ-multiplet is globally well-defined, we study its transformation properties under global symmetries. This analysis leads to several new non-renormalization theorems. We discuss some simple examples in section~3.

In section~4 we consider gauged sigma models and study when a clash with gauge invariance prevents the existence of the FZ-multiplet. This allows us to strengthen some of the non-renormalization theorems of section~2. We then use this stronger form of the non-renormalization theorems to derive a simple identity for the scalar potential of gauged sigma models.

In section~5 we use the fact that many interesting supersymmetric
theories flow to gauged sigma models at low energies to make
general statements about the role of~$D$-terms in models of
SUSY-breaking. We show under very broad assumptions that there can
be no SUSY-breaking vacua (even meta-stable ones) with
parametrically large~$D$-terms. This explains the predominance
of~$F$-term breaking in models of dynamical SUSY-breaking. Our
analysis also sheds new light on the multiplet structure of
Nambu-Goldstone bosons in supersymmetric theories. In particular,
we show that supersymmetry cannot be spontaneously
``shattered''~\shorea.

The results of section~5 immediately raise the question of whether it is possible to build models of dynamical SUSY-breaking with comparable
$D$-terms and $F$-terms. In section~6 we discuss a mechanism which accomplishes this and give three simple, calculable examples. These examples rely on different nonperturbative effects
such as instantons, gaugino condensation, and the emergence of a free magnetic phase. They also show that the significant $D$-terms can be non-abelian; they need not originate from an abelian gauge field. One of our examples is a new calculable limit of the familiar~$3$-$2$ model~\AffleckXZ. In this limit, the model behaves qualitatively differently than in the conventional one. Our mechanism for generating comparable~$D$-terms and~$F$-terms is robust and can easily be implemented in many other models, some of which we
briefly mention. {\it This section is self-contained and can be read independently of the rest of the paper. }

In section~7 we comment on possible future directions, phenomenological applications, and remaining open problems.

\newsec{Supersymmetric Sigma Models}

At low energies, many interesting supersymmetric field theories flow to a weakly-coupled sigma model -- perhaps with IR-free
gauge fields. It is expected that this will happen whenever the
field theory has a strong-coupling scale~$\Lambda$, below which it
is described by massless moduli. In this section we will not
discuss the IR-free gauge fields; this will be done in section~4. We begin
by reviewing some basic facts about supersymmetric sigma
models.

The moduli $\Phi^i$ are described by the Lagrangian
\eqn\smo{{\scr L}=\int d^4\theta \, K\left(\Phi,\b\Phi\right)~,}
where the K\"ahler potential~$K$ is real. In components
\eqn\complag{{\scr L} = - g_{i \jbar} \d^\mu \phi^{*\jbar} \d_\mu \phi^i + ({\rm fermions})~,}
where the K\"ahler metric~$g_{i\jbar}$ on the moduli space~$\CM$ is given by
\eqn\kmet{g_{i \jbar} = \d_i \d_\jbar K~.}
From~\complag\ we see that unitarity requires $g_{i\jbar}$ to be positive definite. The physics of the sigma model is invariant under K\"ahler transformations
\eqn\kt{K\rightarrow K+F\left(\Phi\right)+\bar F\left(\bar \Phi\right)~,}
where $F$ is a holomorphic function of the $\Phi^i$. This invariance is natural from a geometric viewpoint, since K\"ahler transformations are exactly those redefinitions of the K\"ahler potential which leave the metric~$g_{i\jbar}$ invariant.

It is important to note that the K\"ahler potential $K$ need not be a globally well-defined function on~$\CM$. It may differ between different coordinate patches by a K\"ahler transformation~\kt. However, the metric~\kmet\ must be globally well-defined and can be used to construct the  K\"ahler form
\eqn\kform{\Omega = i g_{i \jbar} \, d \Phi^i \wedge d \b \Phi^\jbar~.}
As a consequence of \kmet, the K\"ahler form is closed:~$d \Omega = 0$. In every patch we can thus find a real one-form~$\CA$ such that $\Omega = d \CA$. This $\CA$ is globally well-defined if $\Omega$ vanishes in the cohomology~$H^2(\CM)$.

\subsec{Sigma Models with Global Symmetries}

We now consider a sigma model~\smo\ which is
invariant under a set of continuous global symmetries, and we
derive the associated current multiplets. An infinitesimal global symmetry transformation takes the form
\eqn\inftrans{\delta\Phi^i=\ep\ind X^{(a)i}(\Phi)~,} where the
$X^{(a)i}$ are holomorphic in the fields $\Phi^i$ and
the~$\ep\ind$ are infinitesimal real parameters. The
transformations~\inftrans\ must leave the metric $g_{i\bar j}$
invariant. Geometrically, this means that the~$X^{(a)i}$ are the
components of holomorphic Killing vector fields~$X^{(a)} =
X^{(a)i}\d_i$.

As usual, the symmetry generators $X\ind$ form a Lie algebra
\eqn\algrel{\left[ X\ind, X^{(b)} \right] = - f^{abc} X^{(c)}~, }
where the $f^{abc}$ are real structure constants. Note that the anti-holomorphic complex conjugates $\b X\ind$ satisfy the same algebra~\algrel, and that holomorphic Killing vectors commute with anti-holomorphic Killing vectors: $\left[ X\ind, \b X^{(b)}\right]= 0$.

We will only be interested in symmetry algebras which are direct
sums of compact simple algebras and $U(1)$ algebras, so that we
can choose the structure constants to satisfy the following
properties:

\medskip

\item{1.)} $f^{abc} = 0$ unless all three indices are in the same subalgebra.

\item{2.)} $f^{abc}$ is completely antisymmetric.

\item{3.)} Whenever $a, b$ are in the same compact simple subalgebra, we have
\eqn\km{f^{acd} f^{bcd} = \delta^{ab}~.}

\medskip

\noindent

Let us consider more carefully the invariance of the sigma model under the transformations~\inftrans. While the metric, and therefore the action, are invariant under these transformations, the Lagrangian~\smo\ may pick up a K\"ahler transformation:
\eqn\symkt{\delta {\scr L} = \int d^4\theta \, \ep\ind \left(F\ind + \b F\ind\right)~,}
where the $F\ind$ are holomorphic functions of the $\Phi^i$. This only changes~$\scr L$ by total derivatives, in accordance with the invariance of the action.

We would like to find the conserved currents corresponding to the symmetries~\inftrans. According to the usual Noether procedure, we replace each infinitesimal transformation parameter $\ep\ind$ with an arbitrary chiral superfield $\Lambda\ind$. Now the change in the Lagrangian must take the form
\eqn\extendedchange{\delta {\scr L} = \int d^4 \theta \, \left( \Lambda\ind F\ind + \b \Lambda\ind \b F\ind - i \left(\Lambda\ind -\b\Lambda\ind\right)\CJ\ind\right)~,}
where $\CJ\ind$ is a real superfield. This expression follows from linearity in $\Lambda\ind$ and the requirement that it reduce to \symkt\ when $\Lambda\ind = \ep\ind$, a real constant. Using the explicit form~\inftrans\ of the infinitesimal transformation, we can also write the change in the Lagrangian as
\eqn\lagchange{\delta {\scr L} = \int d^4 \theta \, \left( \Lambda\ind X^{(a)i} \d_i K + \b \Lambda \ind \b X^{(a) \ibar} \d_\ibar K\right)~.}
Since \extendedchange\ and \lagchange\ must agree for every chiral superfield $\Lambda \ind$, we can identify the Noether currents
\eqn\concurr{\CJ\ind = i X^{(a)i} \d_i K - i F\ind~.}
This expression is guaranteed to be real. It is also conserved, as can be checked using the sigma model equations of motion $\b D^2 \d_i K = 0$. Note that the current~\concurr\ is sensitive to the functions $F^{(a)}$ which appear in the K\"ahler transformation~\symkt. This is standard: when the Lagrangian changes by a total derivative due to a symmetry transformation, the Noether current gets modified.

Taking the~$\d_\jbar$-derivative of \concurr, we see that
\eqn\mmeqn{\d_\jbar \CJ\ind = i g_{i \jbar} X^{(a)i}~.}
This is often taken as the defining equation of the currents~$\CJ\ind$.\foot{A real function~$\CJ\ind$ which satisfies~\mmeqn\ is often called the moment map corresponding to the isometry generated by~$X\ind$. The existence of a moment map is locally equivalent to the Killing equation
$$\d_i \left( g_{k \jbar} X^{(a)k} \right) + \d_\jbar \left( g_{i \kbar} \b X^{(a) \kbar}\right) = 0~.$$
} Note that the holomorphic functions $F\ind$ dissapear from~\mmeqn.

\subsec{Consistency Conditions for Global Currents}

We would like to understand whether the current
multiplet~\concurr\ satisfies the consistency conditions~$(2)$ and~$(3)$. We start by examining when a clash with consistency
condition~$(2)$ can arise; conversely, we ask under what conditions~$\CJ\ind$ is globally
well-defined on the moduli space~$\CM$. The key observation is
that~\symkt\ does not define~$F\ind$ uniquely, but only up to a
purely imaginary constant~$i c\ind$. Through~\concurr\ this
corresponds to the freedom of shifting $\CJ\ind$ by a real
constant \BaggerFN: \eqn\constimp{\CJ\ind \rightarrow \CJ\ind +
c\ind~.} This freedom only affects the bottom component $J\ind$ of
the multiplet $\CJ\ind$. In particular, the conserved vector
current $j_\mu^{(a)}$, and hence the charge, is
unaffected. This is a trivial example of an improvement
transformation.\foot{The most general improvement term for a conserved
current~$\CJ$ is of the form $$\delta \CJ = c + D^\alpha \chi_\alpha
+ \b D_\alphadot \b \chi^{\alphadot}~,$$ where~$c$ is a real
constant and~$\chi_\alpha$ is a chiral superfield. In theories like the sigma model, which only contain chiral superfields, the terms involving~$\chi_\alpha$ necessarily contain higher derivatives and we
will not discuss them here.} Note that if the manifold~$\CM$ has
several coordinate patches, then we have the freedom of
shifting~$\CJ\ind$ by a different constant in each
patch.

In this setup, a global obstruction may arise: it might be
impossible to choose the constants~$c\ind$ in such a way that the
current is globally well-defined. More precisely, it is possible for invariant inconsistencies to arise when three coordinate patches intersect (see figure 1). These inconsistencies cannot be removed by shifting the currents. As
usual, a global obstruction of this kind corresponds to a certain non-trivial cohomology class.

\vskip2cm

\epsfxsize=8.5cm \centerline{\epsfbox{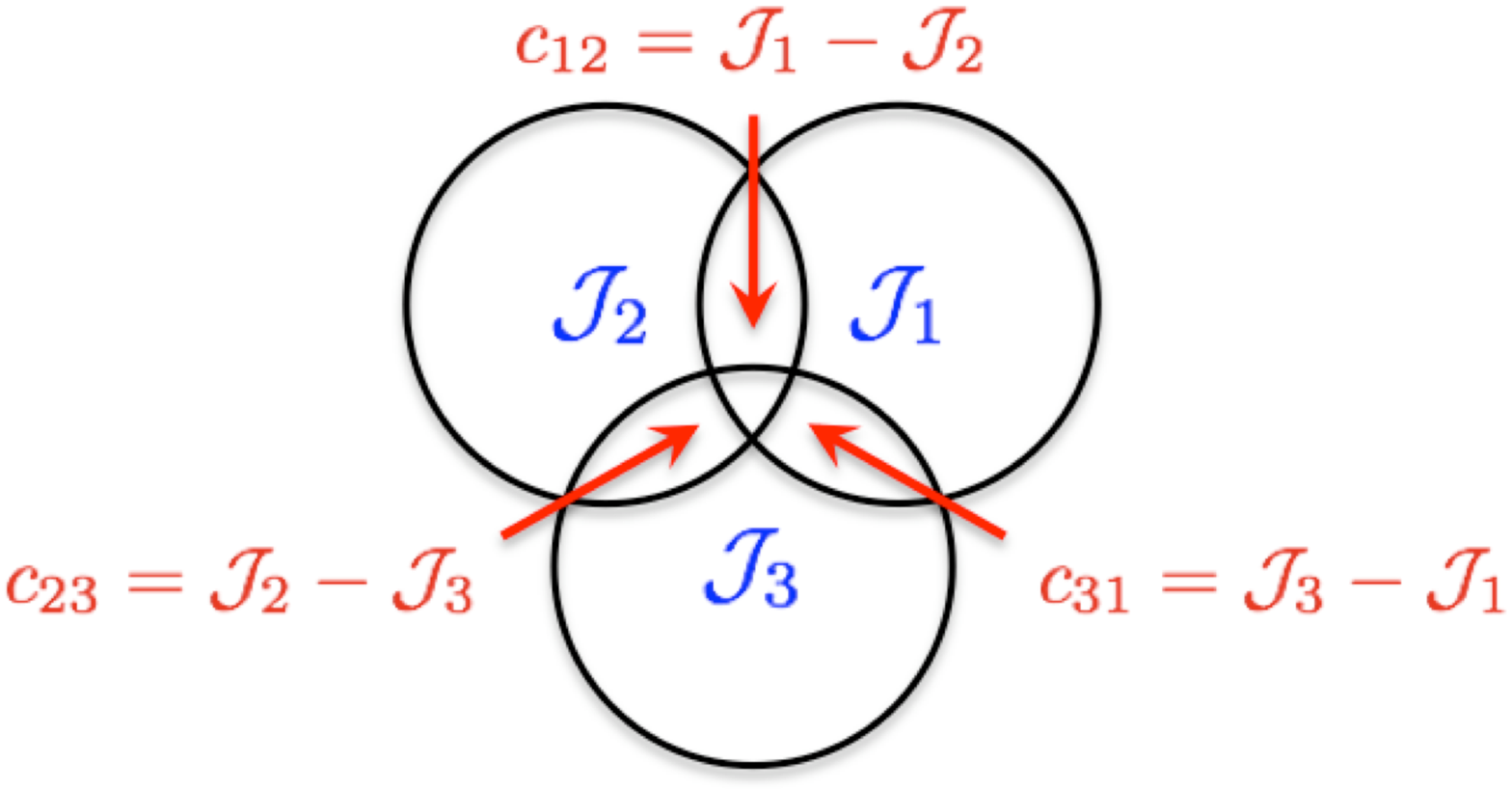}}
\noindent {\bf Figure
1}: In every patch $i = 1,2,3$ the current~$\CJ_i$ is determined
up to a constant. On intersections, the~$\CJ_i$ can differ by
constants~$c_{ij}$. If the invariant sum~$c_{12}+c_{23}+c_{31}$
does not vanish, then the current is not globally
well-defined.

\bigskip

The easiest way to determine this global obstruction is to note
that the right-hand-side of~\mmeqn\ is naturally identified with
the anti-holomorphic part of the real one-form \eqn\oneform{\omega\ind
= -i g_{i\jbar} \b X^{(a)\jbar}d\Phi^i + i g_{i\jbar} X^{(a)i} d
\b \Phi^\jbar~.} Using~\mmeqn, it is easy to check that
$\omega\ind$ is closed. If~$\CJ\ind$ is globally well-defined,
then it is also exact. Conversely, if~$\omega\ind$ is exact then
we can always choose the constants in~\constimp\ to render
$\CJ\ind$ globally well-defined.  Thus, the global symmetry currents~$\CJ\ind$ are
globally well-defined if and only if the one-form~$\omega\ind$ given
by~\oneform\ vanishes in the cohomology~$H^1(\CM)$.\foot{It was
already noted in~\BaggerFN\ that global obstructions to the existence of the~$\CJ\ind$ may arise if the moduli space is not simply-connected.}

It turns out that this obstruction can only arise for abelian groups. For non-abliean groups we can write an explicit expression for
the current multiplet which is manifestly globally well-defined:
\eqn\globalJ{\CJ\ind = f^{abc} \Omega\left(X^{(b)} + \b X^{(b)},
X^{(c)} + \b X^{(c)}\right)~.}
Here~$\Omega$ is the K\"ahler
form~\kform. It is straightforward to check that this~$\CJ\ind$
satisfies~\mmeqn. It can thus be identified with the conserved
current corresponding to the symmetry generated by~$X\ind$. In
section~3 we will see examples of abelian current multiplets which
are not globally well-defined. In section~4 we will explain why
abelian symmetries which lead to such currents cannot be gauged in
the usual minimal way.

We now examine when a clash with consistency condition~$(3)$ can
arise; conversely, we ask when
the~$\CJ\ind$ satisfy the usual current algebra. To understand
this, we consider the transformation~$\delta\ind \CJ^{(b)}$ of the
current~$\CJ^{(b)}$ under the infinitesimal symmetry~\inftrans\
generated by~$X\ind$: \eqn\symvar{\delta\ind \CJ^{(b)} = X^{(a)i}
\d_i \CJ^{(b)} + \b X^{(a)\ibar} \d_\ibar \CJ^{(b)}~.}
Using~\mmeqn, this can be rewritten as \eqn\symvarii{\delta\ind
\CJ^{(b)} =  -\Omega\left( X^{(a)} + \b X^{(a)}, X^{(b)} + \b
X^{(b)} \right)~.}  This implies that $\delta\ind \CJ^{(b)} = -
\delta^{(b)} \CJ\ind$. Taking derivatives on both sides
of~\symvarii\ and using~\algrel\
%\ke,
 we find that
\eqn\curralg{\delta\ind \CJ^{(b)} = - f^{abc} \CJ^{(c)} +
c^{ab}~,} where the~$c^{ab} = -c^{ba}$ are real constants. The
first term on the right-hand-side of~\curralg\ is the one we expect; the antisymmetric constants $c^{ab}$ are
unfamiliar, but play an important role in what follows. These
constants can in principle differ from patch to patch
if~$\CJ^{(a)}$ is not globally well-defined. If the~$c^{ab}$ are non-zero then the full
supersymmetric multiplet~$\CJ\ind$ does not satisfy the usual
current algebra and hence consistency condition~$(3)$ is violated.

As before, these constants are only a problem for abelian currents. They can always be set to zero for non-abelian symmetries by fixing the overall constant in the corresponding~$\CJ\ind$~\refs{\BaggerFN,\HullPQ}. In fact, the expression~\globalJ\ for non-abelian current multiplets leads to vanishing~$c^{ab}$. Non-abelian symmetries thus have globally well-defined current multiplets which obey the usual current algebra.

If $a, b$ both lie in $U(1)$ algebras, then the constants~$c^{ab}$
may be nonzero. Moreover, they cannot be removed by shifting the
currents as in~\constimp. The~$c^{ab}$ correspond to a
well-defined geometric quantity. We can use~\symvarii\ and~\curralg\ to write them in a form which makes this manifest: \eqn\invconst{c^{ab} =
-\Omega\left(X\ind + \b X\ind, X^{(b)} + \b X^{(b)}\right)~.} We
conclude that the abelian~$\CJ\ind$ satisfy the usual current
algebra if and only if the right-hand-side of~\invconst\ vanishes
for all~$U(1)$ indices $a, b$. Note that this condition is local
and thus in general distinct from the condition that the
abelian~$\CJ\ind$ should be globally well-defined. However, the
two conditions are related for compact~$U(1)$
symmetries. If $U(1)_a$ is compact, then~$c^{ab} \neq 0$ only
if~$\CJ^{(b)}$ is not globally well-defined. To see this,
transport~$\CJ^{(b)}$ around a closed loop generated by the
compact symmetry~$U(1)_a$. At each infinitesimal step,~$\CJ^{(b)}$
picks up the constant~$c^{ab}$ and thus cannot come back to
itself: it is not globally well-defined. In section~3 we will see
examples of abelian current multiplets which have
non-zero~$c^{ab}$. In section~4 we will explain why abelian
symmetries which lead to such currents cannot be gauged in the
usual minimal way.

\subsec{Consistency Conditions for the FZ-Multiplet in Sigma Models}

We now discuss the FZ-multiplet for the sigma model~\smo. It is
given by \eqn\sigmaFZ{\CJ_{\alpha\alphadot}^{FZ}=2g_{i\jbar}
D_\alpha\Phi^i\bar D_{\alphadot}\bar\Phi^{\jbar}
-{2\over3}[D_\alpha,\bar D_{\alphadot}]K~,} and satisfies the
conservation equation $\bar
D^{\alphadot}\CJ^{FZ}_{\alpha\alphadot}=D_\alpha X$ with
$X=-{1\over3}\bar D^2 K$. As we discussed around~\kt, K\"ahler
transformations do not change the action. However, we see
from~\sigmaFZ\ that they have an effect on the  FZ-multiplet: under K\"ahler transformations it transforms as
\eqn\imp{\delta
\CJ^{FZ}_{\alpha\alphadot}={2i\over3}\d_{\alpha\alphadot}\left(F-\bar
F\right)~,\qquad \delta X=-{1\over 3}\bar D^2\bar F~.} It can be
checked that~\imp\ only changes the supercurrent and the
energy-momentum tensor by improvement terms. This should be the
case, since K\"ahler transformations do not affect the physics of
the sigma model.

In analogy to the case of global symmetry current multiplets, we now discuss the consistency conditions~$(2)$ and~$(3)$ for the FZ-multiplet. We begin by reviewing the condition under which the FZ-multiplet is globally well-defined on the moduli space~$\CM$.

If we are forced to use multiple patches to cover~$\CM$ and must
perform K\"ahler transformations as we switch from patch to patch,
then~\imp\ implies that a well-defined FZ-multiplet does not exist. To
formulate this condition mathematically, we note that the bosonic
piece of the bottom component of~\sigmaFZ\ is proportional to the
pull-back to spacetime of the one-form~$\CA$ defined after~\kform.
Thus, the FZ-multiplet is globally well-defined if and only if the
one-form $\CA$ is globally well-defined, or in other words when
the K\"ahler form~$\Omega$ is exact~\KomargodskiRB. In particular, a globally well-defined FZ-multiplet does not exist if the moduli space is compact.\foot{On such a space~$\Omega$ cannot be
exact since the volume form of~$\CM$ is a power of~$\Omega$.}

To satisfy consistency condition~$(3)$, the FZ-multiplet should be invariant under the global symmetry
transformations~\inftrans. The danger is that global symmetry
transformations can induce K\"ahler transformations~\symkt\ which
will then lead to a change in the FZ-multiplet. To satisfy
consistency condition~$(3)$ we must therefore require the
holomorphic functions~$F\ind$ which appear in the induced K\"ahler
transformations~\symkt\ to be constants. It is easy to see that these constants must vanish for non-abelian symmetries,
because any non-zero value would break the symmetry explicitly.
This is not necessarily the case for abelian symmetries. The situation is more delicate when these abelian symmetries are gauged; this will be discussed in section~4.

\subsec{Constraints on RG-Flow}

We will now use the results of the previous subsections to state various non-renormalization theorems which constrain the RG-flow of supersymmetric quantum field theories. According to the logic outlined in the introduction, such a non-renormalization theorem holds whenever there can be a clash between certain physical conditions and the existence of particular supersymmetry multiplets, like the global current multiplets~$\CJ\ind$ or the FZ-multiplet. As explained at the beginning of section~2, we consider field theories with a well-defined microscopic formulation in the UV which can be described by a sigma model at low energies.

Assume that the UV theory is invariant under some global symmetries such that corresponding current multiplets~$\CJ\ind$ are globally well-defined and satisfy the usual current algebra.\foot{Here we only discuss global symmetries which are present in the UV theory. We do not consider symmetries which emerge in the IR.} Since these are operator statements, they must hold at every point along the RG-flow. In particular, the discussion in subsection~2.2 implies that the geometry of the moduli space~$\CM$ must be such that all~$U(1)$ current multiplets are globally well-defined, and that the constants~$c^{ab}$ in~\invconst\ vanish for all~$U(1)$ symmetries.

We can actually prove a somewhat stronger result: the constants~$c^{ab}$ are not renormalized along the RG-flow. This can be seen from the fact that these constants appear in the operator equation~\curralg. Alternatively, we can follow 't Hooft and introduce spectator fields to cancel the~$c^{ab}$. Since~\invconst\ implies that non-zero~$c^{ab}$ can only arise for pairs of vector fields which are nowhere vanishing on the moduli space~$\CM$, we can introduce a free chiral spectator field for each such pair and shift its real and imaginary part to set the corresponding~$c^{ab}$ to zero. The symmetry can then be arbitrarily weakly gauged (see section~4), and the non-renormalization theorem follows.

Completely analogously, assume that the theory has a globally well-defined FZ-multiplet in the UV which is invariant under global symmetry
transformations. These conditions must then hold at any point
along the RG-flow. In particular, the discussion of subsection~2.3
implies that the K\"ahler form~$\Omega$ of the moduli space~$\CM$ must be exact, and that the holomorphic
functions~$F\ind$ which appear in the K\"ahler transformations~\symkt\ induced by global symmetry transformations must be constants.
Moreover, these constants must vanish for non-abelian symmetries. In section~4 we will strengthen this result by specifying conditions under which the~$F\ind$ must also vanish for abelian symmetries.

\vskip1cm

\epsfxsize=7cm \centerline{\epsfbox{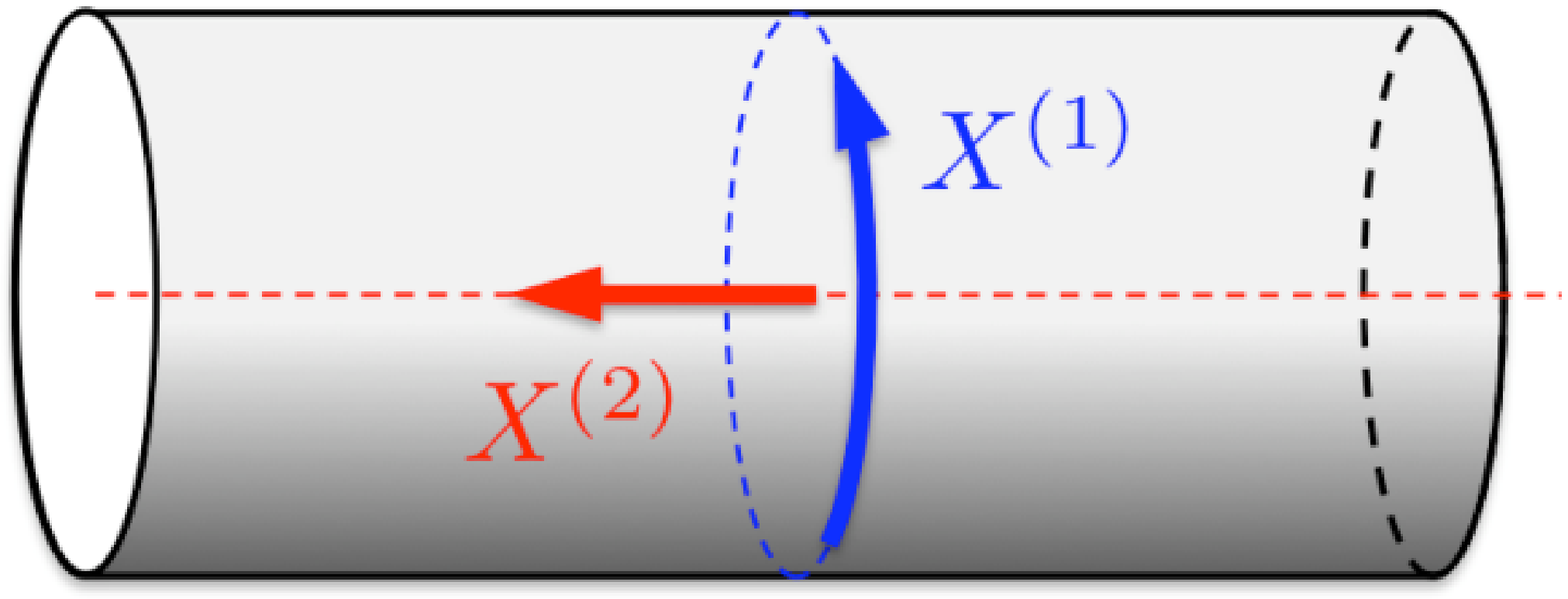}} \centerline{{\bf
Figure 2}: Killing Vectors on the Cylinder}

\newsec{Examples}

In this section we discuss examples which clarify the ideas of section~2.

\subsec{Example 1: Cylinder}

Consider the theory of a single free chiral superfield on a
cylinder:
\eqn\freecyl{\int d^4\theta\,  \Phi^\dagger\Phi~,\qquad
\Phi\sim\Phi+1~.} On this cylinder, we consider the two commuting
Killing vectors \eqn\cylkv{X^{(1)}=-\pa_\Phi~, \qquad
X^{(2)}=-i\pa_\Phi~,} corresponding to shifts in the real and
imaginary part of $\Phi$ respectively (see figure~2). Up to overall additive constants, the corresponding symmetry currents are given by
\eqn\cumucyl{\CJ^{(1)}=i\Phi-i\Phi^\dagger~,\qquad
\CJ^{(2)}=\Phi+\Phi^\dagger~.} We see that $\CJ^{(1)}$ is not
invariant under shifts generated by $X^{(2)}$, and vice-versa.
This means that the constants in~\curralg\ do not vanish:
\eqn\conetwo{c^{12}=-2~.} Hence
consistency condition~$(3)$ is violated. In particular, it is
impossible to gauge the~$U(1) \times U(1)$ symmetry generated
by~$X^{(1)}$ and~$X^{(2)}$ in the usual minimal way (see
section~4).

Another problem of this theory is that the current~$\CJ^{(2)}$ is not globally well-defined so that consistency condition $(2)$ is violated. Upon circling the compact direction of the cylinder,~$\CJ^{(2)}$ does not return to its original value -- it shifts by a constant. This is because the one-form~\oneform\ corresponding to $X^{(2)}$ is $\omega^{(2)} \sim d(\Phi+\Phi^\dagger)$, which is not
exact. Thus, we
cannot even gauge the single~$U(1)$ symmetry generated by $X^{(2)}$ in the usual minimal way (see section~4).

We see that both consistency conditions $(2)$ and $(3)$ are
violated on the cylinder. As was explained in subsection~2.2, a
violation of~$(3)$ necessarily implies a violation of~$(2)$ since
the~$U(1)$ symmetry generated by~$X^{(1)}$ is compact. If we
unwrap the cylinder and consider the same theory on the complex
plane, then the local
obstruction~$c^{12} \neq 0$ remains, even though both currents are
now globally well-defined. In this case, the two conditions are
logically independent since both shift symmetries are non-compact.
Note that the arguments of subsection~2.5 imply that neither the
cylinder nor the plane (with the $U(1)$ symmetries as above) can
arise as moduli spaces of field theories whose current multiplets
satisfy the consistency conditions in the UV.

\subsec{Example 2: $\C\P^1$}

Another instructive example is the $\C\P^1$ model, which is defined by the  K\"ahler potential
\eqn\kapot{K=f_\pi^2\, \log\left(1+\abs{\Phi}^2\right)~.}
Note that in this normalization the chiral superfield $\Phi$ is dimensionless, while $f_\pi$ has dimensions of mass. At least two patches are needed to cover the~$\C\P^1$.  The coordinates in these two patches are related by the inversion~$\Phi\rightarrow 1/\Phi$. This does not
leave the K\"ahler potential invariant, but generates a
K\"ahler transformation \kt\ with $F=-f_\pi^2 \, \log \Phi $. As discussed
in subsection~2.4 this renders the FZ-multiplet globally not well-defined.

An~$SU(2)$ isometry group acts on the~$\C\P^1$ in the usual way. Although these~$SU(2)$
transformations leave the action invariant, some of them induce K\"ahler transformations. The transformation~$\Phi\rightarrow e^{i\alpha}\Phi$ is realized
linearly and leaves the K\"ahler potential invariant, but the other
two~$SU(2)$ transformations induce K\"ahler transformations~\kt\
with~$F$ proportional to~$\Phi$ and~$i\Phi$. This means that the FZ-multiplet is not invariant under the global~$SU(2)$
symmetry. However, since this symmetry is non-abelian, there is no problem with the~$SU(2)$ current multiplets: they are globally well-defined and satisfy the usual current algebra. In
particular, it is possible to gauge this~$SU(2)$ symmetry in the usual minimal way (see section~4).

We see that, like the cylinder, the~$\C\P^1$ model violates consistency conditions~$(2)$ and~$(3)$. The K\"ahler non-invariance of the FZ-multiplet follows from a global obstruction, while the fact that~$SU(2)$ transformations do not leave the FZ-multiplet invariant is a local statement. Like the cylinder, the~$\C\P^1$ model cannot arise as the moduli space of a field theory which satisfies the consistency conditions in the UV.

\newsec{The Gauged Sigma Model}

In the previous sections we have discussed various clashes between the existence of certain current multiplets and the consistency conditions we introduced in the introduction. These clashes led to non-renormalization theorems which we discussed in subsection~2.4. We can say more by gauging the global symmetries of the sigma model. Our discussion will be classical; we will not consider sigma model anomalies.

\subsec{Gauging the Global Symmetries of the Sigma Model}

As always, the gauge transformation corresponding to a global symmetry transformation~\inftrans\ is obtained by replacing the infinitesimal real parameter~$\ep\ind$ by a chiral superfield~$\Lambda\ind$. Under such a gauge transformation the sigma model Lagrangian~\smo\ transforms as in~\extendedchange. To render the theory invariant under such transformations we introduce vector superfields~$V\ind$, so that a gauge transformation with parameter~$\Lambda\ind$ shifts~$V\ind$ by an amount
\eqn\Vgt{\delta V\ind = i\left(\Lambda\ind - \b \Lambda\ind\right) + \cdots~.}
Here the dots represent higher-order terms in~$V\ind$ or~$\Lambda\ind$ which will not be important for us. To leading order in~$V\ind$ we can thus complete~\smo\ to a gauge-invariant Lagrangian by adding a term
\eqn\Lint{{\scr L}' = \int d^4 \theta \, V\ind \CJ \ind~,}
along with the usual minimal kinetic terms for the $V\ind$. Note that the freedom to shift the abelian~$\CJ\ind$ by a constant corresponds to the freedom of adding FI-terms for the abelian gauge fields. The complete Lagrangian ${\scr L} + {\scr L}'$ is now gauge-invariant up to total derivatives:
\eqn\Lgt{\delta\left({\scr L} + {\scr L}'\right) = \int d^4 \theta \, \left( \Lambda\ind F\ind + \b \Lambda \ind \b F\ind\right) + \cdots,}
where the dots represent unimportant higher-order terms and the~$F\ind$ are exactly as in~\symkt. In Wess-Zumino gauge, the~$D$-term scalar potential of the resulting gauge theory is given by
\eqn\Dtermpot{V = \half \sum_a g^2_{a} \left( J\ind\right)^2~.}
Here, the~$g_{a}$ are dimensionless gauge coupling constants and the~$J\ind$ are the bottom components of the global symmetry current multiplets~\concurr\ in the ungauged theory. If the symmetry is linearly realized on the fields in the usual way, this scalar potential reduces to the familiar quartic polynomial in the scalar fields.

We now revisit some of the clashes we investigated in section~2 from the perspective of the gauged sigma model. Since the bottom components~$J\ind$ of the currents~$\CJ\ind$ explicitly appear in the expression~\Dtermpot\ for the scalar potential~$V$, it is clear that we must require the $\CJ\ind$ to be globally well-defined. Moreover, we must require that all~$c^{ab}=0$ so that $V$ is gauge-invariant \HullPQ.

We have already explained in subsection~2.4 that the constants~$F\ind$ must vanish for non-abelian symmetries. In this case the FZ-multiplet is gauge-invariant, since the right-hand-side of~\Lgt\ vanishes.\foot{The FZ-multiplet of the gauged sigma model can be obtained from the FZ-multiplet~\sigmaFZ\ of the ungauged sigma model by replacing
$$K \rightarrow K + V\ind \CJ\ind + \cdots~.$$
A gauge transformation with chiral superfield parameters $\Lambda\ind$ thus changes the FZ-multiplet by
$$\delta \CJ^{FZ}_{\alpha\alphadot} = {2 i \over 3} \d_{\alpha\alphadot} \left( \Lambda\ind F\ind + \b \Lambda \ind \b F\ind\right)~.$$
Therefore, the FZ-multiplet is gauge-invariant if and only if all $F\ind = 0$.} For the abelian case, the arguments of section~2 only allowed us to conclude that the~$F\ind$ are constants. Upon gauging we see that if these constants do not vanish, then the FZ-multiplet is not gauge-invariant. This is expected because in the abelian case these constants are related to FI-terms.

We conclude that all~$F\ind$ appearing in~\symkt\ must vanish in order for the gauged sigma model to to have a gauge-invariant FZ-multiplet. Moreover, the gauged sigma model is only consistent if the symmetry currents~$\CJ\ind$ are globally well-defined and the~$c^{ab}$ in~\invconst\ vanish. This ensures that the scalar potential~$V$ is globally well-defined and gauge-invariant.

\subsec{An Identity for the~$D$-Term Scalar Potential}

We can now prove the following theorem about
the~$D$-term scalar potential~\Dtermpot\ in gauged sigma models: if the
sigma model has a globally well-defined, gauge-invariant
FZ-multiplet, then the scalar potential satisfies the identity
\eqn\thm{V = \half \, g^{\jbar i} \d_i K\,  \d_\jbar V
~.}

To show this, recall from the previous section that the
gauge-invariance of the FZ-multiplet implies that all constants~$F\ind$ in the definition~\concurr\ of the currents $\CJ\ind$
vanish, so that we have \eqn\noFcurr{\CJ\ind = iX^{(a)i} \d_i K~.}
This is guaranteed to be real and globally well-defined. We can
thus write \eqn\identity{V = {i \over 2} \sum_a g_a^2 \, J\ind
X^{(a)i} \d_i K ~.} But using~\mmeqn\ we can express
\eqn\derivpot{\d_\jbar V = i g_{i \jbar} \sum_a g_a^2 \, J\ind
X^{(a)i}~,} so that \identity\ immediately reduces to \thm. In the
next section, we will use~\thm\ to discuss the role
of~$D$-terms in models of SUSY-breaking.

\newsec{$D$-Term SUSY-Breaking}

In this section we use the preceding results to study the role
of~$D$-terms in supersymmetric theories. We will show under very broad assumptions that there can be no
SUSY-breaking vacua (even metastable ones) with parametrically
large~$D$-terms. As a corollary, we also show that supersymmetry cannot be spontaneously ``shattered.''

\subsec{Restrictions on Large~$D$-Terms}

Most known calculable models of
dynamical supersymmetry breaking reduce at very low energies
to a free chiral superfield with linear superpotential. In other
words, the breaking is mostly $F$-term driven. For example, in the
usual treatment of the $3$-$2$ model~\AffleckXZ, the~$SU(3)\times SU(2)$~$D$-terms are set to zero, and there is
a small superpotential on this moduli space of $D$-flat directions
which leads to SUSY-breaking.

We would like to understand whether there are calculable models of dynamical SUSY-breaking which are predominantly~$D$-term driven. A simple model with pure~$D$-term
breaking is the Fayet-Iliopoulos model. However, this model has a FI-term at tree level, and we would like to explicitly forbid such
terms. More precisely, we ask whether~$D$-term
driven SUSY-breaking can occur in theories with an FZ-multiplet in the UV, such as SQCD, chiral theories, quiver gauge theories, and many other examples.

Since we restrict our discussion to calculable models, we only
consider theories in which the low-energy dynamics responsible for
SUSY-breaking can be described by some chiral superfields (often with a superpotential) and possibly some IR-free gauge
fields. In particular, we assume that any strong gauge dynamics
has already been integrated out at a higher scale~$\Lambda$.
Since we are interested in vacua with parametrically
small~$F$-terms, we will set these terms to zero in first
approximation. In section~6 we will discuss models in which this
approximation is not valid; these theories do not give rise to
parametrically small~$F$-terms.

With these assumptions, the low-energy theory reduces to a sigma
model~\smo\ on the moduli space~$\CM$ of~$F$-flat directions. The
IR-free gauge fields are accounted for by weakly gauging some
global symmetries of this sigma model.\foot{In fact, our
discussion also applies to IR-free gauge fields which are not
obtained by gauging a global symmetry; such gauge fields commonly arise in Seiberg duality. We will only need to assume that the gauged sigma model has a globally well-defined, gauge-invariant FZ-multiplet, which is indeed the case in these theories.} If this
weak gauging leads to SUSY-breaking vacua (even metastable ones),
then these putative vacua will have parametrically large~$D$-terms
which are larger than the~$F$-terms by inverse powers of the
(arbitrarily small) IR-free gauge couplings.

We now show that such vacua do
not exist in theories which have a globally well-defined, gauge-invariant
FZ-multiplet in
the UV. In these theories, the low-energy description in terms
of a weakly gauged sigma model also possesses a globally well-defined, gauge-invariant FZ-multiplet. Thus the conditions for the theorem of subsection~4.2 are satisfied, and
the~$D$-term scalar potential satisfies the identity~\thm:
\eqn\thmii{V = \half \, g^{\jbar i} \d_i K\,  \d_\jbar V
~.}
 This identity immediately implies that a critical
point of the potential can only occur when~$V=0$. In other words,
any critical point must be a supersymmetric minimum. Hence, the~$D$-term
scalar potential does not admit SUSY-breaking vacua -- not even meta-stable
ones. This explains the absence of dynamical models with large~$D$-terms. It
does, however, leave open the possibility of models with {\it
comparable}~$D$-terms and~$F$-terms. Such models will be discussed in section~6.

Note that the argument presented above specifically rules out~$D$-terms which are
parametrically larger than the $F$-terms by inverse powers of the gauge couplings. If the theory contains other small numbers (such
as the ratio of charges in~\GregoireJR), then the~$D$-terms can be
enhanced. Since the gauge couplings are IR-free in our setup, it is
natural to approach the problem from the point of view we have taken above. Other small numbers may enter in a model-dependent way and are typically not expected to be present in dynamical models. We therefore do not pursue this possibility further.

It is illuminating to compare our result to a well-known
theorem reviewed in~\BaggerQH. This theorem states that if there is a solution to the~$F$-term equations, then there is a vacuum
that solves both the~$F$-term equations and the~$D$-term
equations; in other words, a SUSY vacuum. This theorem assumes
that the theory has a canonical K\"ahler potential, and that the global symmetries which are being gauged are linearly realized. These assumptions are
important: as we will see below, it is possible to construct sigma models that have vacua
with~$F=0$ and~$D\neq 0$. (Of course, such theories cannot have a well-behaved FZ-multiplet.) Moreover, the theorem makes no statement about the possible existence of metastable SUSY-breaking vacua. Our analysis did not depend on any specific assumptions about the K\"ahler potential or the symmetries. Furthermore, our result rules out all critical points which break SUSY. The key assumption in our analysis is that the low-energy physics can be described by a weakly gauged sigma model with a globally well-defined, gauge-invariant FZ-multiplet. This sigma model can result from complicated, non-perturbative physics at a higher energy scale; it can have a nontrivial K\"ahler potential and global symmetries which are realized in a complicated, nonlinear way.

Let us demonstrate the utility of our result with a simple
example. Consider~$SU(2)$ gauge theory with four chiral superfields~$Q_i$ in the fundamental representation; this model was analyzed
in~\SeibergBZ. The moduli space is parametrized by mesons~$M_{ij}\sim Q_iQ_j$
in the antisymmetric tensor representation of the global~$SU(4)$ symmetry. In the quantum theory, the classical
moduli space~${\rm Pf} (M)=0$ gets deformed to~${\rm Pf} (M)\sim \Lambda^4$. The
K\"ahler potential on this space is not known, and since
the origin is removed, the~$SU(4)$ symmetry must act nonlinearly. We
can gauge the~$SU(4)$ symmetry and study the resulting~$D$-term potential on this interesting deformed moduli space:
\eqn\potential{V= \half \, g_{SU(4)}^2 \sum_{a\in SU(4)}
\left( J\ind\right)^2~.} This theory has a supersymmetric vacuum with unbroken~$SP(4)$ symmetry at
\eqn\susvac{M\sim \Lambda^2\left(\matrix{\sigma^2 &
0\cr 0 & \sigma^2}\right)~.} The question is whether the
potential~\potential\ also admits any metastable SUSY-breaking vacua. (Such vacua could be arbitrarily long-lived if~$g_{SU(4)}$ is sufficiently small.) Since this theory has a globally well-defined, gauge-invariant FZ-multiplet in the UV, our result shows that this cannot happen, even though neither the K\"ahler potential nor the current multiplets~$\CJ\ind$ on the moduli space are known.

Finally, note that models in which a globally well-defined, gauge-invariant FZ-multiplet does not exist can admit vacua with non-vanishing~$D$-terms and vanishing~$F$-terms. A trivial example, which has already been mentioned, is the Fayet-Iliopoulos model. A more interesting example is given by the~$\C\P^1$ model discussed in
subsection~3.2. If we gauge the global~$SU(2)$ symmetry of the~$\C\P^1$, then SUSY
is spontaneously broken on the entire moduli space; the vacuum energy density is a
positive constant proportional to~$g^2 f_\pi^4$, where~$g$ is
the~$SU(2)$ gauge coupling. (The reason this model breaks supersymmetry is described in the next subsection.) This vacuum energy results from pure~$D$-term breaking; all the~$F$-terms vanish.

\subsec{SUSY-Shattering}

 Supersymmetry shattering refers to a particular kind of SUSY-breaking which results in an unusual soft spectrum. When
supersymmetry is unbroken, all matter fields furnish representations of supersymmetry. It is generally expected that when SUSY is spontaneously broken at very low energies (at least in calculable models), we should still be able to group the fields in a way that resembles SUSY multiplets, except that these fields now have SUSY-breaking mass splittings. Put differently, we expect supersymmetry to be broken by the spectrum rather than the field content. However,
sometimes this is not the case. Since this phenomenon is not widely known, let us consider one of the simplest such examples -- the gauged~$\C\P^1$ model from the end of the
previous subsection.

In this theory we gauge the global~$SU(2)$ symmetry which acts nonlinearly
on the~$\C\P^1$. At every point of the moduli space, this~$SU(2)$ symmetry is broken to~$U(1)$. To form two massive supersymmetric vector multiplets, we
need two chiral superfields which can be eaten by the supersymmetric Higgs mechanism. Since there is
only a single chiral superfield, this is impossible and SUSY is broken. Moreover, the two real scalars which are eaten by the ordinary Higgs mechanism to form massive vector fields necessarily correspond to the real and imaginary parts of the complex scalar in the chiral superfield. Hence, the spectrum contains no additional real scalars, which would usually be part of massive supersymmetric vector multiplets. This unfamiliar behavior results from the fact that the bottom component of a single chiral superfield contains two independent Nambu-Goldstone~(NG) bosons. Normally, a NG boson embedded in the bottom component of a chiral superfield is paired with an independent real scalar which is not a NG boson (see \refs{\LercheQA\KugoMA\BandoCC-\BandoFN} for a discussion of NG supermultiplets).

Let us try to generalize this construction to the sigma
model~\smo. We need to understand when the bottom component of a single chiral superfield can contain the NG bosons of two different global
symmetries. Since NG bosons correspond to symmetry transformations, we should view them as Killing vectors on the moduli space. The fact that two NG bosons~$X\ind$ and~$X^{(b)}$ belong to a single chiral superfield is then covariantly expressed by the condition
\eqn\shatgen{\Omega\left(X^{(a)}+\bar X^{(a)},X^{(b)}+\bar
X^{(b)}\right)\neq0~.} When the K\"ahler metric is
flat, this reduces to the condition that the real and imaginary
parts of a single superfield can be identified with independent NG bosons.

Note that~\shatgen\ is precisely the condition we analyzed
in subsection~2.2. For abelian symmetries we saw that~\shatgen\ implies the presence of constants~$c^{ab}$. Thus the theory cannot arise from a conventional UV-completion; moreover the symmetry cannot be gauged in the usual way. Thus both~$a$ and~$b$ should belong to a compact simple Lie algebra. In this case~\globalJ\ implies that there is
a generator~$c$ such that \eqn\vevnonz{\langle
J^{(c)}\rangle\neq0~.}

In order to exploit condition~\shatgen\ to shatter SUSY, we need to gauge
the generators~$a,b$ and thus also~$c$. From~\vevnonz\ we see that the vacuum energy is nonzero and that SUSY is
spontaneously broken by the~$D$-term of the generator~$c$. But we have shown in the previous subsection that
such a vacuum does not exist, if
the theory has a globally well-defined, gauge-invariant FZ-multiplet (we are still assuming that the~$F$-terms vanish).

We therefore conclude that although supersymmetry can be shattered in certain toy models, this cannot be done in interesting dynamical models. Like the result of the previous subsection, this conclusion holds under very broad
assumptions and thus extends
the results of~\refs{\shorea,\shoreb} beyond tree-level models with
canonical K\"ahler potential.

\newsec{Dynamical SUSY-Breaking with Comparable $D$-terms and $F$-terms}

In this section, we present three calculable models which dynamically break supersymmetry with comparable~$D$-terms and~$F$-terms. All of these models have a well-behaved FZ-multiplet; loosely speaking, they saturate the bound~$D \lesssim F$ implied by the results of the previous section. Although our examples rely on different non-perturbative physics to break SUSY, the basic mechanism which leads to comparable~$D$-terms and~$F$-terms is always the same: an~$F$-term potential with runaway directions is stabilized by a~$D$-term potential which results from weakly gauging a global symmetry. The resulting vacuum necessarily breaks supersymmetry with comparable~$D$-terms and~$F$-terms.  This mechanism is robust and can easily be implemented in many other models, some of which we briefly mention below.

\subsec{The 3-2 Model}

The $3$-$2$ model is the simplest model of dynamical SUSY-breaking. We will study this model in an interesting limit of its parameter space in which it is calculable and gives rise to comparable~$D$-terms and~$F$-terms. This limit is different from the conventional one considered in~\AffleckXZ.

The 3-2 model is the unique renormalizable theory with matter
content
\medskip
\eqn\mat{ \matrix{ & [SU(3)] & [SU(2)] & U(1)_Y & U(1)_R
\cr Q^r_A & \square & \square & 1/6 & -1\cr \tilde u_r &
\bar{\square} & 1 & -2/3 & 0 \cr \tilde d_r & \bar{\square} & 1 &
1/3 & 0\cr L^A & 1 & \square & -1/2& 3  }}
\medskip
\noindent Here~$[ \ldots]$ indicates a gauge symmetry. There is a tree-level superpotential given by
\eqn\supthtwo{W=hQ\tilde d L~.}
Conventionally, the model is solved in the limit where the Yukawa coupling~$h$ is the smallest parameter in the problem. In this limit, one focuses on the space of~$SU(3) \times SU(2)$~$D$-flat directions and minimizes the~$F$-term scalar potential on this space. In this approach, the~$D$-terms are bound to be negligible (see~\SeibergQJ\ for a detailed discussion). SUSY-breaking is triggered by the interplay of two terms in the~$F$-term potential: a tree-level contribution from~\supthtwo\ and a non-perturbative piece.

We would like to study the~$3$-$2$ model in a different limit in which the~$SU(2)$ gauge coupling~$g_2$ is the smallest parameter in the problem, so that we can set it to zero in first approximation. Now the dynamics and the low-energy behavior will be very different than in the conventional limit. Fortunately, the model is still calculable.

We will take the quickest path to obtaining the vacuum structure of this model. Note that~$L^A$ is a classical modulus. We assume that the SUSY-breaking vacuum is at large values of~$L^A$ (this will be justified by self-consistency). To simplify the analysis, we take the vev of~$L^1$ to be large and set the vev of~$L^2$ to zero, at the expense of manifest~$SU(2)$ invariance. On this flat direction the quarks~$Q_1$,~$\tilde d$ have a large mass~$hL^1$ and can be integrated out. The remaining low-energy theory consists of an~$SU(3)$ gauge theory with quarks~$Q_2, \t u$ so that~$N_f = 1$, and the massless singlets~$L^1, L^2$.
In terms of the original~$SU(3)$ strong-coupling scale~$\Lambda$, the strong-coupling scale~$\Lambda'$ of the low-energy theory is given by~$\left(\Lambda'\right)^{8} = h L^1 \Lambda^7$. Quantum effects generate a non-perturbative superpotential given by~\refs{\DavisMZ,\AffleckMK}:
\eqn\ADS{W=2\, {\left(\Lambda'\right)^4\over \sqrt{Q_2\tilde
u}}=2h^{1/2} \Lambda^{7/2}\sqrt{L^1\over Q_2\tilde u}~.}
Up to gauge transformations, the~$SU(3)$~$D$-flatness conditions imply that $Q_2=\tilde u={1\over \sqrt2}(a,0,0)$ with $a \in \C$. Since the K\"ahler potential is canonical at large field values, we can effectively describe the theory with three massless degrees of freedom:
\eqn\effthe{K_{\rm eff}=a^\dagger a+\left(L^1\right)^\dagger L^1+\left(L^2\right)^\dagger L^2~,\qquad
W_{\rm eff}= 2\sqrt2 h^{1/2} \Lambda^{7/2} \, {\sqrt{L^1} \over a}~.}

This theory has a runaway as~$a \rightarrow \infty$. In this regime, $SU(3)$ is spontaneously broken to~$SU(2)$ and the non-perturbative superpotential~\ADS\ is a result of gaugino condensation in the unbroken~$SU(2)$ gauge theory. This is different from the conventional limit of the $3$-$2$ model, where the gauge symmetry is completely broken and the non-perturbative superpotential is generated by an instanton.

To stabilize the runaway, we weakly gauge the $SU(2)$ symmetry.
The full scalar potential is then given by\foot{For consistency,
we continue to use structure constants which are normalized to
satisfy~\km, even though this differs from the usual convention by
a factor of~$\sqrt 2$.}
\eqn\scapot{V=2h\Lambda^7\left({1\over |L^1| |a|^2}+{4|L^1|\over
|a|^4}\right)+{g_2^2\over 16}\left(|L^1|^2+\half|a|^2\right)^2~.}
Note that this only works because the light degrees of freedom on
the runaway are charged under~$SU(2)$ and contribute to
the~$D$-terms with equal sign. To leading order in~$g_2$, the
location of the vacuum is given by \eqn\sol{L^1\approx0.69\,
\Lambda h^{1/7} g_2^{-2/7}~,\qquad a\approx 2.27 \, \Lambda
h^{1/7}g_2^{-2/7}~.} For small~$g_2$ these vevs are parametrically
large, so that our analysis is self-consistent. SUSY is broken and
the vacuum energy density scales like~$h^{4/7}g_2^{6/7}\Lambda^4$;
it receives comparable contributions from the~$F$-terms and from
the non-abelian~$SU(2)$~$D$-terms. In this vacuum, the Goldstino
is essentially an equal mixture of the matter fermions and
the~$SU(2)$ gauginos. In contrast, the field vevs and the vacuum
energy in the conventional treatment of the~$3$-$2$ model are
independent of~$g_2$. The dominant contributions to the vacuum
energy and the Goldstino come from~$F$-terms and matter
fermions respectively.

In our analysis of this model, we have taken the quickest path to obtain the leading order answers~\sol. These results receive corrections which are suppressed by powers of~$g_2$, which should be studied systematically. (The same comment applies to the other models in this section.) In addition, it would be interesting to understand whether it is possible to interpolate between our limit and the conventional limit of the~$3$-$2$ model in a continuous, calculable way. The~$3$-$2$ model can be generalized in several different ways. This leads to large classes of theories which rely on essentially the same SUSY-breaking mechanism (see
for instance~\IntriligatorFK\ and the reviews~\refs{\PoppitzVD,\ShadmiJY}). We expect many of these models to admit new calculable SUSY-breaking vacua of the kind we just described for the~$3$-$2$ model.

\subsec{The 4-1 Model}

We repeat the analysis of the previous subsection for the~$4$-$1$ model~\refs{\DineAG,\PoppitzFH} to give an example of a theory
with large abelian~$D$-terms. Here we will no longer keep track of numerical coefficients.

The model has gauge group~$SU(4) \times U(1)$ and matter content
\medskip
\eqn\content{\matrix{ & [SU(4)] & [U(1)] & U(1)_R \cr S & 1 & 4 &
6 \cr F_i & \square & -3& 0 \cr \tilde F^i & \b{\square} & -1 & -4 \cr A_{ij} & \square\square &
2& 0}}
\medskip
\noindent There is also a tree-level superpotential \eqn\super{W=h\tilde
F^iF_iS~.}
\noindent Like before, we are interested in the limit in which the~$U(1)$ gauge coupling~$g_1$ is the smallest parameter in the problem. This limit of the~$4$-$1$ model was studied in~\CarpenterTZ. Our
analysis only differs in that it focuses on the light degrees of freedom on the moduli space. See also~\ElvangGK\ for some discussion of departing from $U(1)$ flatness.

When we go to large values of the classical modulus~$S$, the~$SU(4)$
quarks~$F_i, \t F^i$ have a large mass~$h S$ and can be integrated out.
Below this scale we have an~$SU(4)$ gauge theory with an
anti-symmetric tensor field. In terms of the original~$SU(4)$ strong-coupling scale~$\Lambda$, the strong-coupling scale~$\Lambda'$ of this new theory is given by~$\left(\Lambda'\right)^{11} \sim h S \Lambda^{10}$. Up to gauge transformations,  the~$SU(4)$~$D$-flatness conditions imply that
\eqn\solD{A \sim a \pmatrix{ \sigma^2&0\cr 0&\sigma^2}, \qquad a \in \C~.}
This breaks the gauge symmetry to~$SP(4)$ at a scale~$\sim a$. The remaining pure Yang-Mills theory has beta function~$9$ so that its strong-coupling scale~$\Lambda''$ is given by $(\Lambda')^{11} \sim a^2 (\Lambda'')^9$. Gaugino condensation in the~$SP(4)$ theory generates a superpotential
\eqn\fouronegcont{W \sim \left(\Lambda''\right)^3 \sim \left({hS\Lambda^{10} \over a^2 }\right)^{1/3}~.}
The K\"ahler potential is canonical at large field values.

As before, the theory has a runaway as~$a \rightarrow \infty$ which can be stabilized by weakly gauging the~$U(1)$ symmetry; this works because~$S$ and~$a$ both have positive~$U(1)$ charge. The full scalar potential then takes the form
\eqn\effpotefo{V \sim h^{2/3}\Lambda^{20/3}\left({1\over
|S|^{4/3} |a|^{4/3}}+{|S|^{2/3}\over |a|^{10/3}}\right)
+ g_1^2 \left(|S|^2+ |a|^2\right)^2~.}
\noindent At leading order in~$g_1$ the vacuum is given by
\eqn\minimum{S\sim a \sim \Lambda h^{1/10}
g_1^{-3/10}~.}
SUSY is broken and the vacuum energy scales like $h^{2/5} g_1^{4/5} \Lambda^4$; it receives comparable contributions from the~$F$-terms and from the abelian~$D$-term.

Like the~$3$-$2$ model, the~$4$-$1$ model has many generalizations, some of which
are reviewed in~\refs{\PoppitzVD,\ShadmiJY}.

\subsec{In the Free Magnetic Phase}

In this subsection we discuss a simple generalization of the~$3$-$2$ model which also breaks SUSY by using the weak gauging of a global symmetry to stabilize a runaway direction. Depending on the parameters, the non-perturbative effect responsible for the runaway behavior is either gaugino condensation or the emergence of a free magnetic phase. Like in the previous subsection, we will not keep track of numerical coefficients.

The theory is simply the~$3$-$2$ model with two extra flavors~$\Psi_i,\tilde \Psi^i \; (i = 1,2)$ of vector-like~$SU(3)$ quarks (we suppress their color indices), so that the matter content of the model is given by
\medskip
\eqn\matconfm{ \matrix{ & [SU(3)] & [SU(2)] \cr Q^r_A & \square &
\square \cr \tilde u_r & \bar{\square} & 1  \cr \tilde d_r &
\bar{\square} & 1 \cr L^A & 1 & \square \cr \Psi_{1,2}\left(\tilde
\Psi^{1,2}\right) & \square \, \left(\bar{\square}\right) & 1 }}
\medskip
\noindent We give the extra quarks~$\Psi, \tilde\Psi$ a small mass~$m \ll \Lambda$, where~$\Lambda$ is the~$SU(3)$ strong-coupling scale. This means that the~$SU(3)$ gauge theory has~$N_f = 4$ light flavors. The tree-level superpotential is given by
\eqn\supelem{W = hQ\tilde d L+m\Psi_i\tilde \Psi^i~.}

Like before, we will assume that the~$SU(2)$ gauge coupling~$g_2$ is very small (below, we will exactly specify how small); for simplicity we take~$h \sim 1$. Like in the previous examples, the theory has a runaway when~$g_2 = 0$. Turning on~$g_2$ stabilizes the vacuum at a point in field space which approaches the origin as we increase~$g_2$.

Even though~$m \ll \Lambda$, the location of the SUSY-breaking vacuum is still calculable. (This analysis closely parallels the discussion in subsection~6.1.) Very far out on the classical moduli space parametrized by~$L^1$ and~$Q_2=\tilde u \sim (a,0,0)$, the~$SU(3)$ symmetry is spontaneously broken to~$SU(2)$. In this~$SU(2)$ gauge theory, the quarks~$\Psi,\tilde \Psi$ are heavy and can be integrated out; the resulting pure Yang-Mills theory undergoes gaugino condensation (with a calculable coefficient). Like before, a non-perturbative potential arises from the moduli dependence of the condensate. The resulting low-energy theory is essentially the same as the one in subsection~6.1; only the effective strong-coupling scale is different because the massive quarks~$\Psi, \t \Psi$ have been integrated out. The vacuum is located at
\eqn\expvalues{L^1\sim a\sim m^{2/7}\Lambda^{5/7}h^{1/7}g_2^{-2/7}~.}
SUSY is broken and the vacuum energy scales like $m^{8/7}\Lambda^{20/7}h^{4/7}g_2^{6/7}$. Again, the contribution from~$D$-terms and~$F$-terms is comparable. This solution can only be trusted if the expectation values~\expvalues\ are above the strong-coupling scale~$\Lambda$. This is the case as long as~$g_2\ll m/\Lambda$.

It is interesting to see what happens when we increase the gauge coupling beyond this point, so that~${m\over \Lambda}\ll g_2\ll 1$. Even though the vacuum now lies in the strong-coupling region, the theory is still calculable thanks to the dual magnetic description~\SeibergPQ.

The magnetic dual of~$SU(3)$ SQCD with~$N_f = 4$ light flavors consists of a~$4\times 4$ meson matrix~${M_{i}}^j$, four flavors of magnetic quarks~$q^i,\tilde
q_i$ and no IR-free gauge fields. Decomposing the meson matrix into~$2 \times 2$ blocks,
\eqn\mesmat{M=\left(\matrix{{X_i}^j &
{Y_i}^j \cr \tilde {Y_i}^j &
{Z_i}^j } \right) = { 1 \over \Lambda} \left(\matrix{\Psi_i\tilde \Psi^j &
\Psi_i\, (\tilde d, \tilde u)\cr Q_i \tilde \Psi^j & Q_i \, (\tilde d, \tilde u) } \right)  ~,} the superpotential of the dual theory can be written as
\eqn\supmesmat{W_{d}=q^i {M_i}^j \tilde
q_j+h \Lambda {Z_i}^1 L^i-\mu^2\left({X_1}^1+{X_2}^2\right)~,}
where~$\mu^2\sim m\Lambda$.\foot{For clarity of presentation, we are not explicitly introducing the usual magnetic Yukawa coupling, which we take to be~$\CO(1)$. When we are computing loops in this IR-free coupling, the expansion parameter is the loop factor~$1 \over 16 \pi^2$.} We also
decompose the magnetic quarks as $q=(\chi^1,\chi^2,\rho^1,\rho^2)$,
and similarly for the~$\tilde q$.

As in~\IntriligatorDD, this theory breaks SUSY at tree level by the rank
condition: the~$F$-term equations for~${X_{i}}^j$ take the form~$\chi_i \tilde \chi^j = \mu^2{\delta_i}^j$. This
equation cannot be satisfied because the rank of the left-hand-side is at most one. However, unlike the situation in~\IntriligatorDD, not all mesons acquire~$F$-term vevs. The mesons~${Z_i}^j$ are somewhat decoupled from the SUSY-breaking fields~${X_i}^j$. This is the case because some of the quarks in the original electric theory~\supelem\ are massless.

The dynamics of the dual theory~\supmesmat\ has been analyzed in~\refs{\GiveonWP,\GiveonNE} (see~\AmaritiUZ\ for generalizations); here we will just state the results. At tree-level, the theory has a number of massless pseudomoduli. Since the couplings are IR-free, we can calculate loop corrections to the pseudomoduli potential.  At one-loop, all pseudomoduli except the~${Z_j}^2$ are stabilized at the origin. Because they are somewhat decoupled from the SUSY-breaking fields, the~${Z_j}^2$ remain massless at one-loop.

We denote one of the~${Z_j}^2$ by $Z$ (as before, we are picking a preferred~$SU(2)$ direction). An effective potential for~$Z$ is generated at two loops:
\eqn\effpotZ{V_{\rm eff} \sim {1\over (16\pi^2)^2} \Biggl \{ \matrix{-\mu^2 \left| Z\right|^2+\cdots & \left| Z \right| \ll\mu \cr -\mu^4\left(\log {\left|Z\right|^2\over\mu^2}\right)^2+\cdots &  \mu\ll \left| Z \right| \ll\Lambda }}
In both parameter regimes, the dots represent terms which are suppressed. Beyond the scale~$\Lambda$, we cannot use the magnetic description to determine the potential. We see from~\effpotZ\ that the magnetic theory has a runaway for large~$Z$. In the UV, this runaway is completed by the runaway we discussed above in the electric theory.

As in our previous examples, the runaway in the magnetic theory can be stabilized by reintroducing the gauge coupling~$g_2$.  Let us first consider the regime~$\mu\ll |Z| \ll\Lambda$, where the full effective potential for~$Z$ is now given by
\eqn\minprobone{V_{\rm eff} \sim -{\mu^4\over \left(16\pi^2\right)^2}\left(\log {|Z|^2\over\mu^2}\right)^2+g^2_2|Z|^4~.}
This fixes~$Z\sim {\mu\over\sqrt\delta}\left(\log{1\over\delta}\right)^{1/4}$, where $\delta=16\pi^2g_2$. Note that the vacuum energy~$\sim \mu^2$ receives its dominant contribution from the tree-level~$F$-term of the~${X_i}^j$ mesons. In this regime, the ~$D$-term contribution to the vacuum energy is suppressed because the magnetic theory has a new small parameter: the IR-free Yukawa coupling in which we perform the loop expansion.

For this solution to be reliable, we must require that~$|Z| \ll\Lambda$, or equivalently~$g_2\gg {\mu^2\over \Lambda^2}$. Since~$\mu^2\sim m\Lambda$, we see that~$g_2\gg {m\over\Lambda}$. Note that both the electric and the magnetic descriptions break down when~$g_2\sim m/\Lambda$. This fact seems to depend non-trivially on non-holomorphic data; we view it as a consistency check of the duality.

If we increase~$g_2$ even further (roughly,~$g_2$ needs to be larger than a loop factor), then the vacuum eventually enters the regime~$|Z| \ll\mu$ where the effective potential is given by the first line of~\effpotZ. Stabilizing this against the quartic~$D$-term potential, we find that~$Z \sim {\mu\over  16\pi^2 g_2}$. As before, the vacuum energy~$\sim \mu^2$ is dominated by the tree-level contribution.

\newsec{Open Problems}

In this section we briefly mention some open problems which we have not addressed above, as well as some other ideas which might be interesting to pursue. We hope to report on these topics in the future.

\medskip

$\bullet$ We have described situations in which the global current multiplets~$\CJ\ind$ can fail to exist as well-behaved operators. Analogously to~\KomargodskiRB, there should be other (perhaps larger) multiplets which contain the global symmetry currents and do satisfy the consistency conditions. It should then be understood how to gauge these multiplets.

$\bullet$ We have shown that any theory with a well-behaved
FZ-multiplet cannot break supersymmetry with parametrically
large~$D$-terms (even in a metastable vacuum). General statements about
the scalar potential of supersymmetric theories are also familiar in
other contexts. For instance, the $F$-term potential in theories
with canonical K\"ahler potential is known to have pseudo-flat
directions~\RayWK. Some general results have also been obtained for certain~$\CN=2$ theories~\AntoniadisNJ, and weakly deformed~$\CN=2$
theories~\refs{\OoguriIU\MarsanoMT-\AuzziKV}. It is an interesting and important problem to find more classes of theories for which the space of SUSY-breaking vacua can be studied in a controlled and general way.

$\bullet$ We have shown that it is impossible to shatter SUSY in models which have a well-behaved FZ-multiplet in the UV. However, we have not constructed a renormalizable theory where SUSY is shattered, for instance with a FI-term. The $\C\P^1$-example discussed above can easily be UV-completed to a renormalizable model with a FI-term which shatters SUSY
classically. However, the minimal such completion is anomalous quantum mechanically. As was pointed out in~\IntriligatorBE, consistent models which shatter SUSY might be interesting for Higgsed gauge mediation~\refs{\GorbatovQA\IntriligatorFR-\BuicanVV}.

$\bullet$ We have constructed several models which break
supersymmetry with comparable~$D$-terms and~$F$-terms. In these examples, a runaway direction is stabilized by weakly gauging a global symmetry. This opens several directions for further study. First, it would be interesting to implement this mechanism in theories with deformed moduli spaces and compare the resulting models to~\refs{\IzawaPK,\IntriligatorPU}. Second, we have not discussed theories in the conformal phase. It is not understood how to analyze such theories or how to study their putative
SUSY-breaking vacua. Finally, models with significant~$D$-terms
have a variety of phenomenological applications. For example, they
can be used to cover the parameter space of gauge
mediation~\refs{\MeadeWD,\BuicanWS}. Under certain assumptions, it
was argued in~\DumitrescuHA\ that this cannot be done without using~$D$-terms. Moreover,~$D$-terms may play an important role in constructing models in which the SUSY-breaking vacuum is stable, but nevertheless leads to a nonzero gaugino mass at leading order~\matos. Significant~$D$-terms are also useful
in the context of anomaly-mediated SUSY-breaking and in models with Dirac gauginos; see, for example~\refs{\CarpenterTZ,\GregoireJR,\BenakliPG,\BenakliGI}.

$\bullet$ Our analysis has lead to general statements about the~$D$-term scalar potential. This may prove useful in studying the space of exactly marginal deformations of four-dimensional superconformal theories; this space is usually called the conformal manifold. As explained
in~\GreenDA, the RG-flow near a fixed point on the conformal manifold is a gradient flow
controlled by a~$D$-term potential on the space of coupling constants. Our results can be used to understand global properties of the conformal manifold, if the gradient flow persists beyond the regime discussed in~\GreenDA.

\bigskip
\bigskip
\noindent {\bf Acknowledgments:}

We would like to thank M.~Buican, A.~Giveon, D.~Green,
K.~Intriligator, M.~Rocek, N.~Seiberg, and B.~Wecht for comments.
The work of TD was supported in part by NSF grant PHY-0756966 and
a Centennial Fellowship from Princeton University. The work of ZK
was supported in part by NSF grant PHY-0503584.  MS was supported
by World Premier International Research Center Initiative (WPI
Initiative), MEXT, Japan. Any opinions, findings, and conclusions
or recommendations expressed in this material are those of the
authors and do not necessarily reflect the views of the National
Science Foundation.

\listrefs
\end